\PassOptionsToPackage{hyphens}{url}

\documentclass[preprint,12pt]{elsarticle}

\usepackage{amssymb}   % Provides various useful mathematical symbols
\usepackage{amsmath}   % Provides advanced environments for equations

\usepackage{float}     % Provides the [H] option to "force" float placement
\usepackage{siunitx}   % For professional SI units (e.g., \SI{10}{\micro\meter})
\usepackage{threeparttable}

\usepackage[margin=1in]{geometry} 
\usepackage{booktabs}  % Highly recommended for professional-quality tables (replaces \hline)
\usepackage{xcolor}
\usepackage[final]{changes}

\usepackage{hyperref} 
\journal{Journal of Manufacturing Processes}

\begin{document}

\begin{frontmatter}



%\title{In-situ monitoring of directed energy deposition using fringe projection profilometry}

\title{Layer-Wise Anomaly Detection in Directed Energy Deposition using High-Fidelity Fringe Projection Profilometry}


\author[label1]{Guanzhong Hu} %% Author name
\author[label1]{Wenpan Li}
\author[label1]{Rujing Zha}
\author[label1]{Ping Guo\corref{cor1}}
\ead{ping.guo@northwestern.edu}
\cortext[cor1]{Corresponding author}

\affiliation[label1]{organization={Department of Mechanical Engineering, Northwestern University},%Department and Organization
            city={Evanston},
            state={IL},
            country={USA}}

%% Abstract
\begin{abstract}

Directed energy deposition (DED), a metal additive manufacturing process, is highly susceptible to process-induced defects such as geometric deviations, lack of fusion, and poor surface finish. This work presents a build-height-synchronized fringe projection system for in-situ, layer-wise surface reconstruction of laser-DED components, achieving a reconstruction accuracy of $\pm\SI{46}{\micro\meter}$. From the reconstructed 3D morphology, two complementary geometry-based point cloud metrics are introduced: local point density, which highlights poor surface finish, and normal-change rate, which identifies lack-of-fusion features. These methods enable automated, annotation-free identification of common deposition anomalies directly from reconstructed surfaces, without the need for manual labeling. By directly linking geometric deviation to defect formation, the approach enables precise anomaly localization and advances the feasibility of closed-loop process control. This work establishes fringe projection as a practical tool for micrometer-scale monitoring in DED, bridging the gap between process signatures and part geometry for certifiable additive manufacturing.
\end{abstract}

%     \vspace{1in}
%     \centering
%     \includegraphics[width=\textwidth]{Graphical_abstract_white.png}

% \begin{highlights}
%  \item In-situ and layer-wise monitoring of micrometer-scale 3D morphology during directed energy deposition using a build-height–synchronized fringe projection system.

%  \item Geometry-based anomaly detection that identifies under-/over-deposition and lack-of-fusion features without manual annotation.

%  \item High-fidelity defect localization through controlled experiments, achieving sub-$\SI{50}{\micro\meter}$ accuracy and advancing toward closed-loop quality control in additive manufacturing.

%  \end{highlights}

%% Keywords

\end{frontmatter}



\section{Introduction}
\label{intro}
%% Labels are used to cross-reference an item using \ref command.

\emergencystretch=1em

Directed energy deposition (DED) is a versatile additive manufacturing (AM) process widely employed for the fabrication, repair, and modification of metallic components~\cite{svetlizky_directed_2021}. In powder-based laser-DED, metal feedstock is delivered into a laser-induced molten pool, forming a series of beads that accumulate layer by layer to produce a three-dimensional (3D) part~\cite{fan_partially_2023}. This layer-wise material addition enables DED to offer several advantages over traditional manufacturing methods, including the ability to produce complex topologies~\cite{chen_directed_2025}, integrate multiple materials~\cite{feenstra_critical_2021}, and achieve region-specific flexibility~\cite{kim_selective_2021}. Despite its advantages, the inherent complexity of DED process, encompassing heat transfer~\cite{sun_numerical_2020}, material flow~\cite{webster_situ_2024}, and solidification, raises concerns regarding inevitable defect formation~\cite{zhang_numerical_2021}. Common anomalies include height deviations, lack of fusion (LOF), spatter particles, and gas pores, which are mainly observable on the newly deposited layer before the subsequent layer is added~\cite{imran_advancements_2024}. These anomalies are typically manifested through deviations in measurable quality indicators such as layer height, surface roughness, and geometric morphology, which can be quantitatively assessed during the build. Compared to the more mature metal AM process, powder bed fusion (PBF), anomalies in DED are both more frequent and more detrimental due to its distinctive deposition characteristics. Height deviations are particularly critical~\cite{li_single-sensor_2024}, since the absence of a recoater makes layer height entirely dependent on process parameters, causing cumulative bulging or valleys that can degrade dimensional accuracy and even interrupt the build. LOF is also more severe in DED~\cite{dos_santos_paes_lack_2022}, as powder blown into the melt pool leads to inconsistent track overlap, and the relatively thick bead geometry produces large, elongated pores compared to the smaller-scale defects typical of PBF layers of $\sim\SIrange{20}{50}{\micro\meter}$. Finally, surface finish is generally poorer in DED, as it often exhibits waviness and powder adhesion~\cite{liu_review_2021}. To detect anomalies in a timely manner before they are obscured by subsequent layers, which may compromise mechanical strength and structural integrity, in-situ and interlayer monitoring becomes critical for quality assurance.

Two-dimensional (2D) image acquisition and processing remain central to DED research. One study reported a representative setup in which a CMOS camera was positioned laterally to the melt pool and translated synchronously with the laser nozzle to capture continuous grayscale intensity images of single-track deposits~\cite{lu_online_2025}. This configuration enabled dynamic monitoring of deposition shrinkage, but became ineffective once multiple tracks consolidated into a continuous surface. For defect monitoring in DED, the extreme processing environment and the requirements for real-time and scalable inspection make visual methods particularly appealing. Although 2D image-based approaches have been explored to observe the shape of DED-printed components~\cite{liu_review_2021}, the complex surface texture of DED-printed parts, combined with shadowing effects and specular reflections,  limits their reliability in detecting subtle or recessed regions. To achieve more definitive and reliable detection, accurate 3D morphology measurements are required.

Thus, several 3D measurement techniques have been employed for in-situ morphology monitoring of DED. One widely adopted technique is line laser scanning (LLS), which captures the surface morphology by traversing the layer with a focused laser beam~\cite{binega_online_2022}. It is a robust method suitable for deployment in manufacturing inspections under extreme conditions~\cite{11080679}. This method provides high-resolution surface profiling and enables the evaluation of surface quality~\cite{wang_-situ_2025} as well as estimation of melt pool depth when used in conjunction with melt thermometric measurements~\cite{jeon_online_2021}. However, it typically requires multiple scans and subsequent registrations to reconstruct the full field~\cite{ghanadi_review_2024}, which is time-consuming and can introduce errors during registration post-processing, ultimately affecting measurement accuracy. Digital image correlation (DIC) has been employed for surface morphology monitoring by tracking natural or artificially applied speckle patterns on the object surface to infer 3D shape changes over time~\cite{wang_situ_2023}. This approach enables deformation monitoring and facilitates the reconstruction of the complete thermo-mechanical history during the build process~\cite{haley_-situ_2021}. However, its effectiveness deteriorates significantly on feature-deficient surfaces and under varying lighting conditions, thereby limiting its reliability in practical DED applications. Closed-range photogrammetry (2D Imaging) reconstructs surface contours from camera imagery~\cite{ning_height_2024}, which is capable of capturing lateral profiles and two-dimensional deposition geometry. However, it generally lacks the depth resolution required for accurate surface characterization and is not well suited for monitoring the dynamic, layer-by-layer evolution inherent to DED processes.

The limitations of these existing methods emphasize the need for an active illumination, full-field 3D measurement system that offers high depth resolution, low cost, and easy integration into existing DED setups. To address this, we propose an anomaly detection technique based on fringe projection profilometry (FPP). FPP is a structured light method capable of micrometer-level 3D reconstructions of object surfaces by projecting fringe patterns and analyzing their deformations using phase demodulation techniques. A summary and comparison of these in-situ surface metrology techniques are presented in Table~\ref{tab:techniques_comparison}. Unlike thermal or conventional optical monitoring approaches, FPP directly quantifies surface morphology and offers high spatial resolution over a large field of view~\cite{10807740}. This shift from proxy-based monitoring to true geometric evidence highlights that the morphology itself encodes the presence of process anomalies, thereby enabling more reliable and automated defect detection.

\begin{table}[H]
\centering
\footnotesize %\footnotesize
\begin{threeparttable}
\caption{Comparison of in-situ surface metrology techniques for DED.}
\label{tab:techniques_comparison}

\begin{tabular}{
>{\raggedright\arraybackslash}p{3cm}
>{\raggedright\arraybackslash}p{2.4cm}
>{\raggedright\arraybackslash}p{4cm}
>{\raggedright\arraybackslash}p{2.2cm}
>{\raggedright\arraybackslash}p{2cm}
}\toprule

\textbf{Method} & \textbf{Measurement Principle} & \textbf{Limitations in DED Inspection} & \textbf{Axial Resolution} & \textbf{Lateral Resolution} \\
\midrule
Laser Line Scanning~\cite{binega_online_2022}\cite{wang_-situ_2025}\cite{wang_situ_2023} & Line-by-line scanning & Slow acquisition (non-full-field), high-cost & \SIrange{1.5}{247}{\micro\metre}\tnote{a} & \SI{12}{\micro\metre}\\

\addlinespace

Digital Image Correlation~\cite{wang_situ_2023} & Full-field 3D via stereo-DIC & Low accuracy on low-texture surfaces & \SI{12}{\micro\metre} & \SIrange{2}{3}{\micro\metre}\\
\addlinespace

2D Imaging~\cite{ning_height_2024} & 2D image capture & Lacks 3D topographical data & N/A & \SI{250}{\micro\metre}\tnote{b}\\

\addlinespace

FPP (this work) & Full-field 3D structured light & Sensitive to specular reflections & \SI{46}{\micro\metre} & \SI{12}{\micro\metre} \\
\bottomrule

\end{tabular}

\begin{tablenotes}
\footnotesize
\item[a] Varies with system configuration.
\item[b] Limited by pixel size and FOV.
\end{tablenotes}

\end{threeparttable}
\end{table}

Although commercial FPP systems have been successfully implemented in PBF for build surface topography characterization, geometric feature extraction~\cite{zhang_situ_2016}, full-bed inspection after powder recoating~\cite{liu_-situ_2020}, and prediction of local density variations~\cite{remani_-situ_2024}, their deployment in DED remains largely unexplored due to the substantially different build strategies and defect morphologies inherent to the process. In PBF, the powder bed is lowered after each layer so the measurement surface remains at a fixed height, keeping the optics in focus~\cite{dev_singh_powder_2021}. In contrast, DED adds material on a stationary substrate, so the surface of interest rises continuously~\cite{ahn_directed_2021}. This necessitates dynamic refocusing to preserve triangulation geometry and measurement fidelity across layers. To address this, we integrate a build-height–synchronized FPP module that repositions its focal plane according to the nominal layer height, keeping both projector and camera within their optimal working ranges throughout deposition. This module can directly measure the as-built 3D morphology at micrometer resolution during the build, thereby effectively closing the gap between real-time process monitoring and final geometry validation. The resulting surface point clouds then serve as the foundation for quantitative quality evaluation and automated anomaly detection.

Conventional non-learning-based image processing techniques, though widely applied in PBF for tasks such as melt pool monitoring and defect detection, exhibit limited robustness when transferred to DED. This limitation arises primarily from the unstable surface brightness of DED builds, which is strongly influenced by illumination variability, specular reflection, and the inherent geometric complexity of deposited features. Moreover, the analysis of 3D morphology data in DED remains nontrivial: accurate localization of surface irregularities often requires manual annotation, introducing subjectivity and limiting scalability. To overcome these challenges, we introduce a geometry-driven point-cloud analysis pipeline that leverages local point-distribution density and the normal-change rate (NCR) as complementary descriptors for unsupervised anomaly detection and segmentation. Point-distribution density is sensitive to sampling sparsity caused by under-deposition or surface discontinuities, whereas NCR highlights abrupt curvature transitions typically associated with LOF or surface collapse. By jointly exploiting these geometric signatures, the proposed approach reduces reliance on manual intervention, enhances detection consistency, and provides a scalable framework for quantitative surface quality assessment in DED.

\begin{figure}[H]
    \centering
    \includegraphics[width=0.95\linewidth]{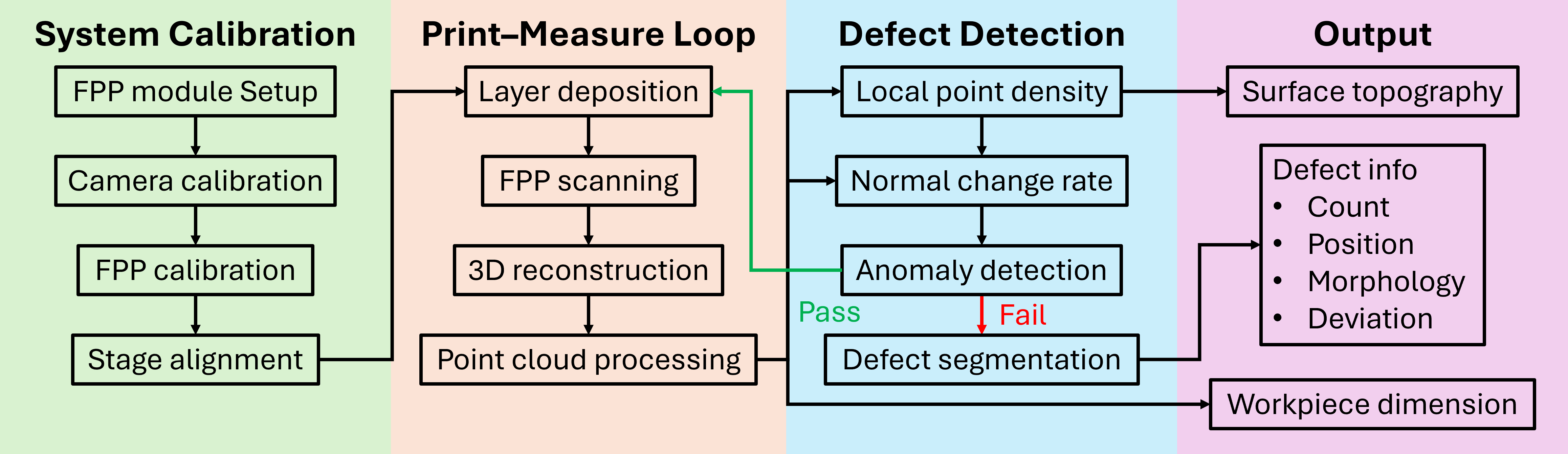}
    \caption{Workflow of the in-situ DED monitoring framework.}
    \label{fig:System_workflow}
\end{figure}

Together, these elements form a scalable, high-resolution, and cost-effective framework for in-situ, interlayer surface morphology monitoring and unsupervised anomaly detection in DED (Figure~\ref{fig:System_workflow}), advancing the instrumentation and analytics needed for in-process quality assurance. Compared with focus-variation and laser confocal microscopy, our framework achieves comparable accuracy without interrupting DED production; it further outperforms DIC by eliminating additional lighting or surface treatment requirements~\cite{hsu_vision-based_2019}. Quantitatively, our method achieves a vertical root mean square error (RMSE) of $\pm\SI{13}{\micro\meter}$ on calibration blocks and $\pm\SI{46}{\micro\meter}$ on actual DED-fabricated surfaces, and demonstrates a repeatable lateral resolution of $\pm\SI{11.8}{\micro\meter}$, which is more consistent than that of laser line scanners under practical working distances~\cite{suresh_recent_2025}. Beyond reconstruction, the geometry-driven anomaly-detection pipeline provides the capability to automatically localize and segment surface irregularities, reducing manual effort and improving consistency in quality assessment.

\section{Methodology}
\subsection{Fringe Projection Profilometry}
FPP operates by projecting a sequence of sinusoidal fringe patterns onto the object surface~\cite{zuo_phase_2018}. When these patterns are captured from an angle by a camera, surface height variations cause local shifts in the observed fringe phase relative to a flat reference plane. Through phase demodulation, these shifts are expressed as a wrapped phase map \(\phi_w\). After unwrapping the phase, this yields a continuous phase distribution \(\phi\) proportional to the object’s height via the system’s triangulation geometry~\cite{feng_calibration_2021}. In practice, precisely determining the relative pose between the projector, camera, and the object surface is challenging, particularly when the imaging sensors are enclosed within the device housing~\cite{10934722}. Fortunately, the system is equipped with a translation mechanism that allows the projector and camera assembly to move along the optical axis. This ensures that the relative position between the surface and the imaging system remains constant during each measurement. As a result, the complex geometric relationships can be effectively represented by a single phase-to-height calibration constant \(C\)~\cite{zhang_high-speed_2019}.

In this study, a conventional phase-shifting algorithm is employed to retrieve the unwrapped phase \(\phi\), which is linearly related to the surface height \(z\) as:
\begin{equation}
z = C \phi,
\label{eq:height_phase}
\end{equation}
where \(C\) is the phase-to-height calibration constant~\cite{odowd_effects_2021}. This constant is obtained experimentally by correlating phase changes with the known height profile of a reference surface. The calibrated FPP system is subsequently integrated into the DED process, in which measurement fidelity must be maintained throughout fabrication as the build height increases dynamically.

\subsection{System Calibration}
\label{System_Calibration}

Accurate system calibration is essential for achieving high-fidelity 3D reconstruction and precise anomaly localization. In this work, the calibration procedure consisted of two major steps: camera parameter calibration and phase-to-height calibration. The intrinsic parameters of the camera were estimated using MATLAB’s Camera Calibration Toolbox with a standard checkerboard target. Multiple images captured at varying orientations and positions were processed to determine the focal lengths, principal point coordinates, and lens distortion coefficients. These parameters were subsequently applied for distortion compensation, ensuring geometric consistency and preserving measurement accuracy in the reconstructed surface data.

Lateral pixel resolution was determined using the known dimensions of a 1.5 mm-pitch checkerboard placed on the build plate within the sealed DED chamber. The physical spacing, combined with the observed number of pixels per checker square, provided the pixel-to-length scaling factor for subsequent 3D measurements. At a calibrated working distance of 205 mm, the camera’s field of view was approximately 49 mm $\times$ 36 mm. Given the image resolution of 4128 $\times$ 3008 pixels, this corresponded to an effective lateral resolution of 11.8 µm/pixel.

\begin{figure}[H]
    \centering
    \includegraphics[width=0.48\linewidth]{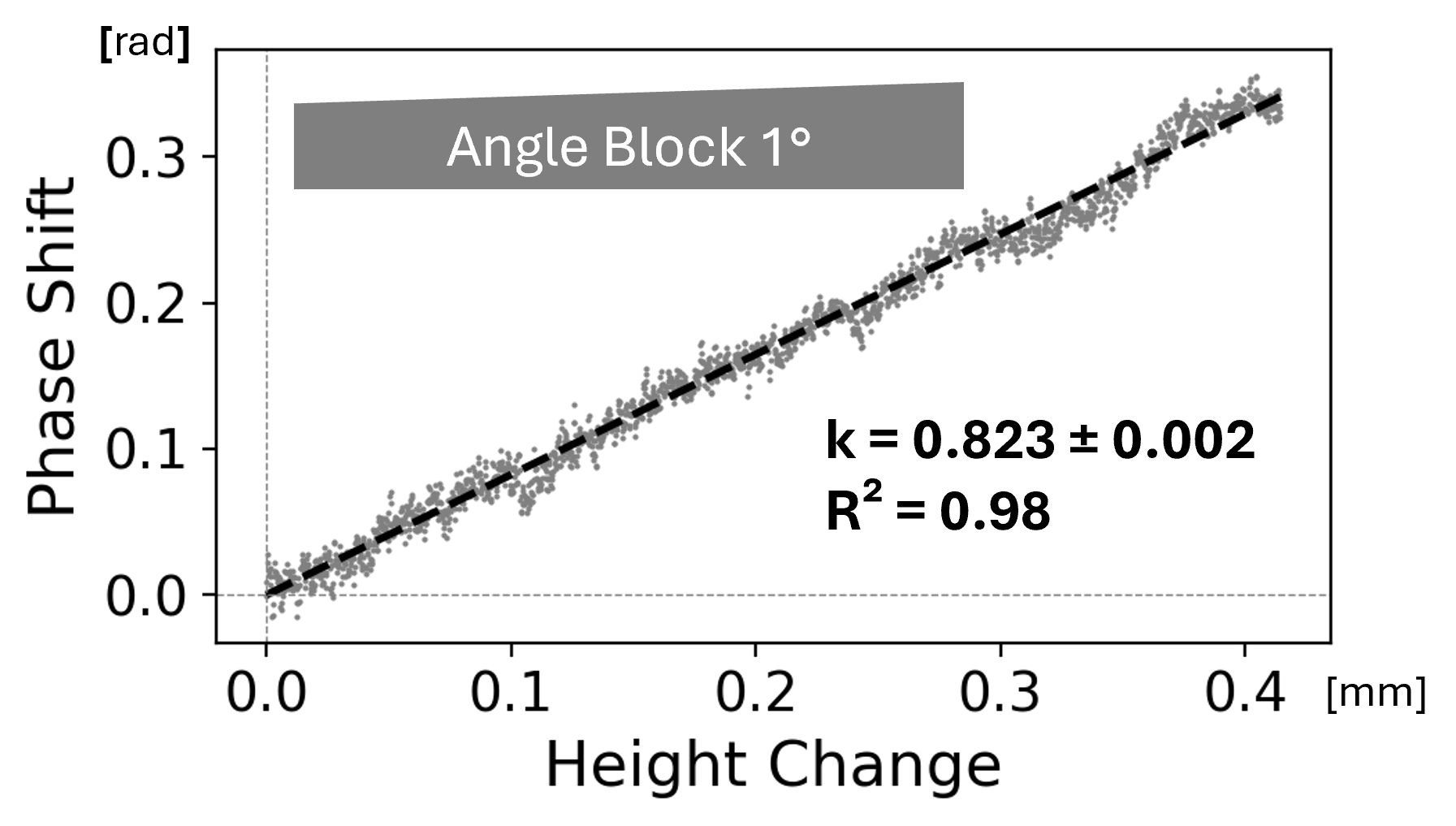}
    \caption{Phase-to-height calibration using a $1^\circ$ angle block. A linear fit yields the calibration constant $C = 0.823$.}
    \label{fig:Angle_block}
\end{figure}

After the camera parameter calibration, phase-to-height mapping was subsequently performed to enable accurate full-field 3D surface reconstruction. A linear phase-to-height calibration was conducted using an angle gauge block with a known $1^\circ$ inclined surface was placed on the build plate. Based on the calibrated lateral resolution, a 20~mm segment along the slope was selected as the calibration region. As shown in Figure~\ref{fig:Angle_block}, the unwrapped phase values within the segment were correlated with the analytically determined height profile from the known inclination angle and measured length of the reference surface. A least-squares linear fit yielded a calibration constant $C = 0.823$, indicating the height change per unit phase difference. 

\subsection{System Validation}
\label{System_Validation}

To establish baseline confidence in the absolute measurement capability of the FPP system, a two-stage validation was carried out.

\begin{figure}[H]
    \centering
    \includegraphics[width=0.48\linewidth]{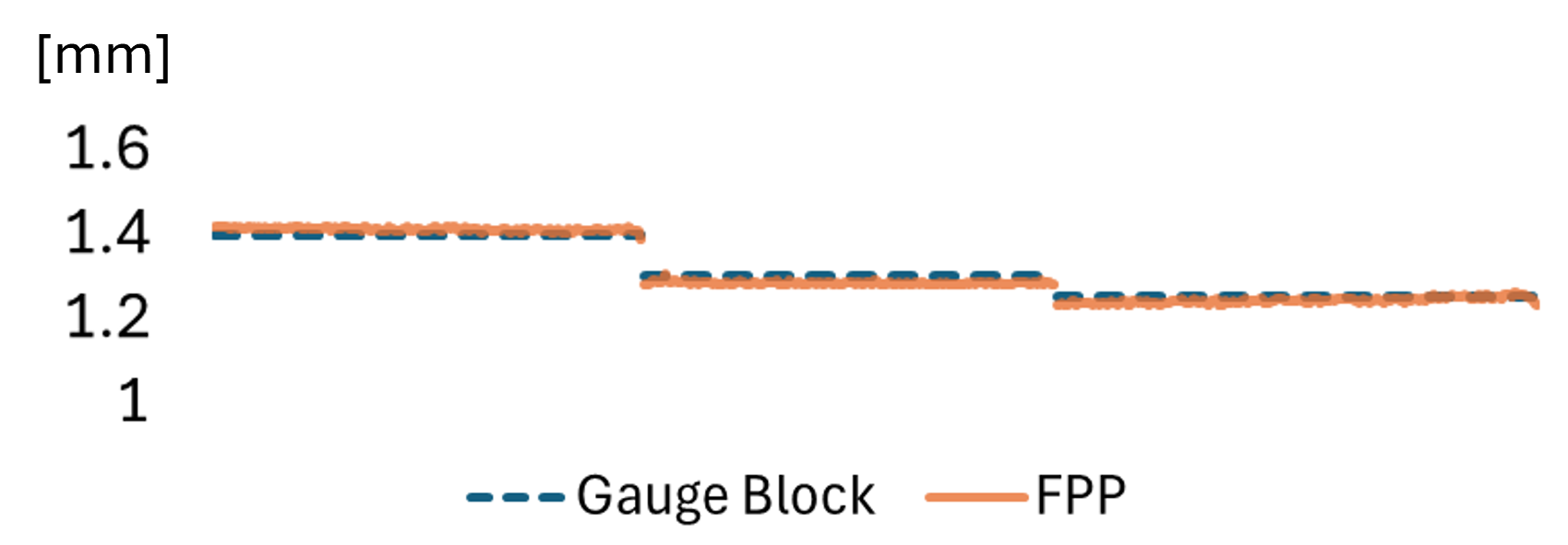}
    \caption{Comparison between FPP and gauge block}
    \label{fig:Gauge_block}
\end{figure}

First, a baseline assessment was performed using gauge blocks (516-946-26, Mitutoyo, Japan). Three blocks of different heights were arranged in a staircase configuration, and the measured step heights were compared with the nominal values. The FPP system achieved a root mean square error (RMSE) of $\SI{13}{\micro\meter}$, as shown in Figure~\ref{fig:Gauge_block}, thereby confirming its fundamental accuracy.

Building on this baseline validation, a more comprehensive evaluation was conducted to quantitatively assess the system's performance in reconstructing mesoscale surface morphology of DED components. For this purpose, measurements obtained by the FPP system were compared against a commercial focus-variation microscope (InfiniteFocus G4, Bruker Alicona, Austria), which served as the reference standard. 

\begin{figure}[H]
    \centering
    \includegraphics[width=0.48\linewidth]{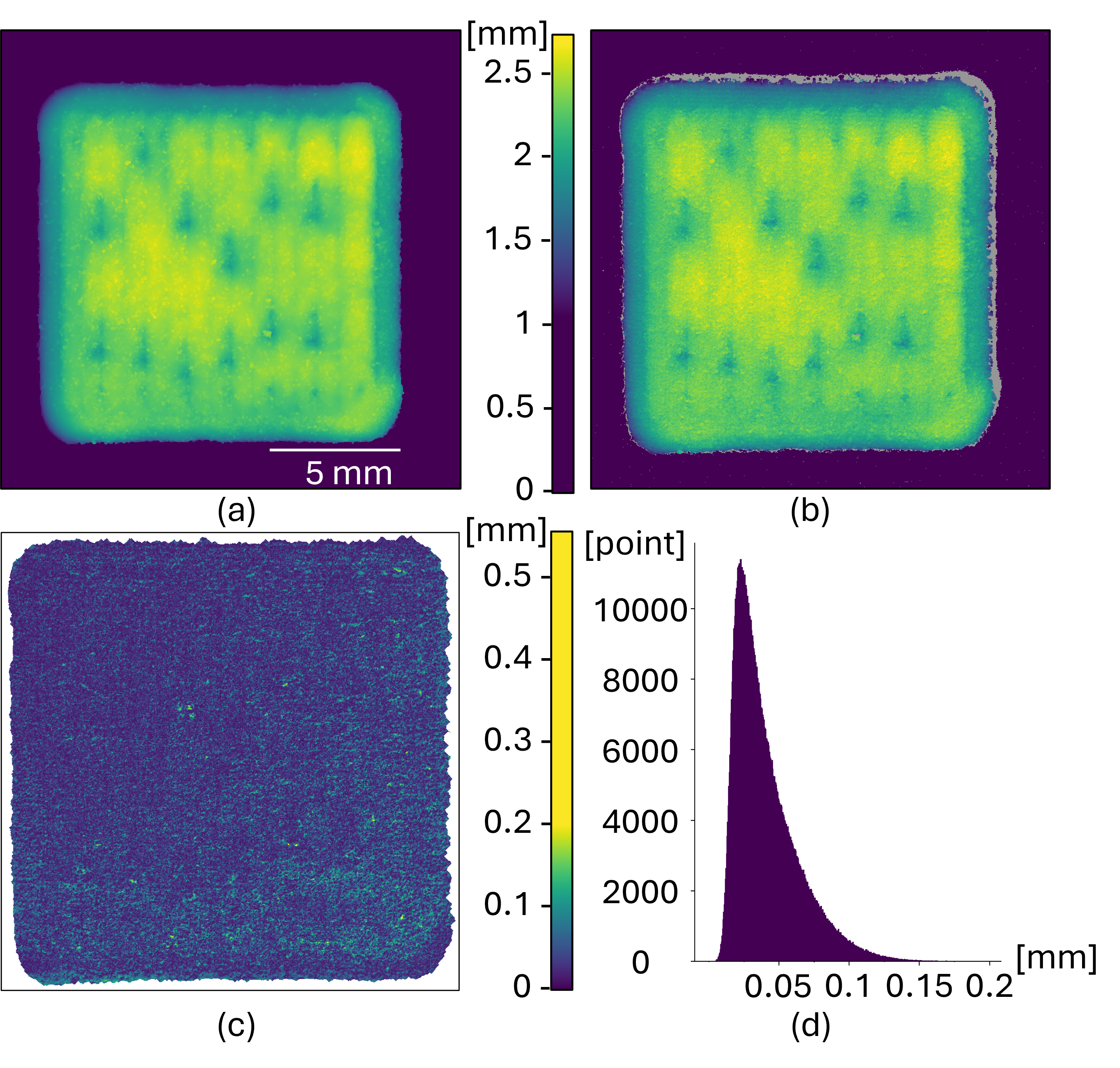}
    \caption{Comparison between FPP and focus-variation system measurements: (a) height map reconstructed from FPP; (b) corresponding height map obtained via 3D microscope; (c) pixel-wise deviation magnitude between the two reconstructions; and (d) statistical histogram of deviations.}
    \label{fig:FPP_microscope_comparison}
\end{figure}

To evaluate the influence of surface gradient on reconstruction accuracy, two surface conditions were analyzed. The first specimen region contained 12 intentionally introduced depressions, each with a depth of approximately $0.5\,\mathrm{mm}$, exhibiting pronounced local gradients. The corresponding height maps obtained by the FPP system and the 3D microscope are presented in Figure~\ref{fig:FPP_microscope_comparison}(a) and \ref{fig:FPP_microscope_comparison}(b). Pixel-wise deviations between the two reconstructions were computed as the Euclidean distance between corresponding height values within the overlapping measurement domain, and the results are visualized in Figure~\ref{fig:FPP_microscope_comparison}(c). The statistical distribution of these deviations is presented in Figure~\ref{fig:FPP_microscope_comparison}(d), based on $1{,}037{,}505$ valid data points binned into $500$ equally spaced intervals over the $0$--$0.2\,\mathrm{mm}$ range. Key results include:
\begin{itemize}
    \item $72.84\%$ of data points lie within $\pm 0.05\,\mathrm{mm}$,
    \item $97.68\%$ of data points lie within $\pm 0.10\,\mathrm{mm}$,
\end{itemize}

Subsequently, a nominal as-built DED surface was evaluated to assess performance on regions with smaller slopes and smoother transitions. For this nominal surface, the reconstruction accuracy improved to an RMSE of $\SI{43}{\micro\meter}$. Quantitative analysis indicates a higher fidelity, with the proportion of points within $\pm 0.05\,\mathrm{mm}$ increasing to $78.74\%$, and those within $\pm 0.10\,\mathrm{mm}$ rising to $99.20\%$.

These results indicate that reconstruction accuracy is dependent on local surface gradients and roughness. For reference, on standard calibration blocks with a nearly Lambertian finish and minimal slope variation, the system achieves a significantly lower RMSE of $\SI{13}{\micro\meter}$. The increase in RMSE to \SIrange{43}{46}{\micro\meter} on actual DED components is thus attributed to the non-ideal optical properties of the metal surface, specifically local shadowing effects and reduced fringe contrast in areas with steep slopes. Despite these challenges, the system reliably maintains a lateral resolution of $\SI{11.8}{\micro\meter}$/pixel, confirming its suitability for capturing mesoscale irregularities in additive manufacturing.

\section{System Integration}

\subsection{Experimental Setup}
\label{Experimental_Setup}
The experimental setup integrates three primary components: (1) an FPP system for interlayer surface morphology measurement during DED process; (2) a motion stage enabling repositioning of the FPP module along a single axis; and (3) a powder-fed laser-based DED system for sample fabrication.

The FPP module supplied by Phase3D Inc. consists of a projection unit and an imaging unit. The projection system utilizes a DLP-based projector (LC3010-RGB10/OF, Keynote Photonics, USA) to project high-contrast RGB fringe patterns onto the build surface for structured light 3D measurement. The imaging unit employs a CMOS camera (Alvium 1800 U-1242m, Allied Vision, Germany) equipped with a 50 mm focal length lens (86574, Edmund Optics, USA). The native resolutions of the projector and the camera were 1280 $\times$ 720 and 4128 $\times$ 3008 pixels, respectively.

During the experiments, measurements were conducted within a sealed chamber equipped with a laser safety observation window (ACRX-BB2, Kentek, USA), which provides broadband optical attenuation and exhibited particularly high rejection in the blue spectral range (200--532 nm, optical density (OD) 6+, i.e., transmittance below $10^{-6}$). Blue-channel projection was selected not only due to the shorter wavelength~\cite{liang_short_2014}, which improved phase sensitivity and signal-to-noise ratio, but also because the window effectively suppressed external blue-light interference, thereby enhancing measurement stability.

Both the projector and the camera were rigidly mounted to ensure mechanical stability and preserve their relative spatial calibration throughout the experiment. The camera was oriented perpendicularly to the region of interest (the build plate), while the projector was positioned at a 15$^\circ$ angle relative to the camera's optical axis. The optimal focus distance for both the camera and the projector was set to 205 mm from the build plate along the surface normal.

\begin{figure}[H]
\centering
\includegraphics[width=0.48\linewidth]{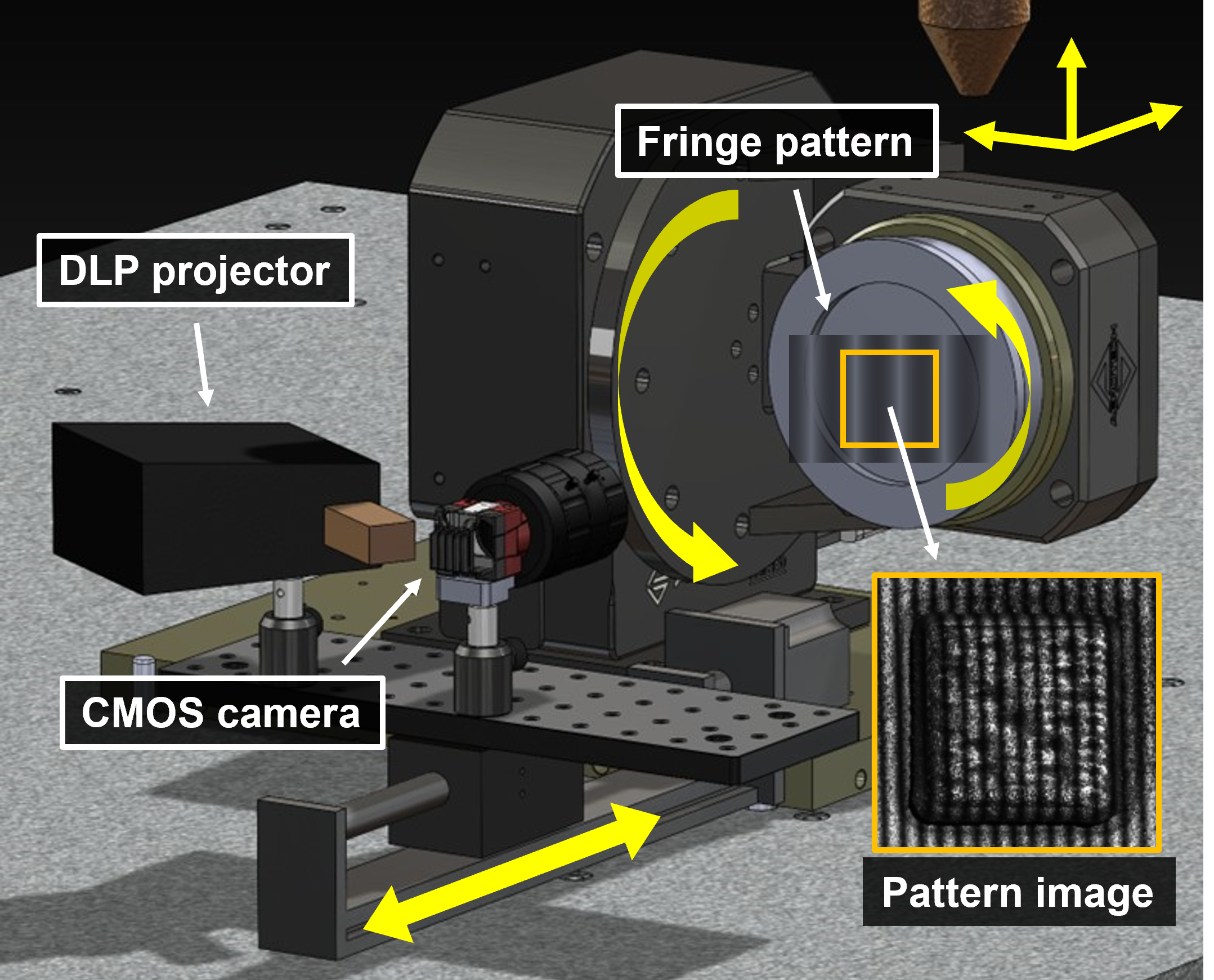}
\caption{Rendering of the experimental setup.}
\label{fig:Setup_Rendering}
\end{figure}

This configuration ensures that the projected fringe pattern fully covers the build plate within the camera’s field of view, which spans approximately 49 mm $\times$ 36 mm, sufficient to capture the entirety of the printed sample (12 mm $\times$ 12 mm). The setup rendering of the measurement module is illustrated in Figure~\ref{fig:Setup_Rendering}.

Accurate surface metrology during the DED process requires consistent imaging geometry throughout the build. Since the processing chamber is sealed to minimize contamination and reconditioning needs, the FPP system was designed for remote operation. It was mounted on a precision linear actuator (FSK40F200-10C7, FUYU, China) and controlled by an Arduino Uno Rev3. 

Because of limited vertical clearance above the build plate, the FPP module could not be installed overhead. Instead, the system was positioned laterally, and the build plate was rotated after each layer to face the system. In this arrangement, the increase in build height translates into a horizontal shift in the FPP frame of reference (Figure~\ref{fig:Setup_Rendering}).

To maintain a fixed optical baseline, the lateral position of the FPP module was adjusted in step with the build height:
\begin{equation}
X_m(n) = X_0 + h_n = X_0 + n \cdot h_{\text{layer}},
\label{eq:actuator_motion}
\end{equation}
where $X_0$ is the initial lateral offset, $n$ is the layer index, and $h_{\text{layer}}$ is the nominal layer thickness. This ensured the module remained within its calibrated measurement volume. The actuator's resolution of \SI{50}{\micro\meter} enabled precise positioning to satisfy this constraint across all layers.

Experiments were conducted using the Additive Rapid Prototyping Instrument (ARPI) system at Northwestern University~\cite{zha_-situ_2024}. The build chamber features a coaxial laser--powder nozzle equipped with gas-assisted delivery and integrated with a fiber-coupled laser. A multi-axis rotary stage at the base of the chamber holds the circular substrate, allowing both rotation about its vertical axis and tilt adjustment for surface reorientation, as illustrated in Figure~\ref{fig:Setup_Rendering}. The nozzle provides three translational degrees of freedom, and in combination with the rotary stage, the system offers full five-axis spatial control.

Key system components include a 1000~W, 1070~nm continuous-wave fiber laser (YLR-1000, IPG Photonics, USA), providing the primary energy source for material deposition. Motion control is achieved using Aerotech linear and rotary stages (USA). The deposition optics consist of a Precitec optical column (Germany) integrated with a Fraunhofer ILT COAX8 nozzle (Germany) for coaxial powder injection. Powder feed is supplied through a PowderMotionLabs X2 precision feeder (USA), with flow stabilized by 99.999\% purity argon gas (O$_2$ $<$ 1~ppb) serving as both carrier and shielding atmosphere. The build substrate is a 15.8~mm-thick 1018 low-carbon steel disc (\#7786T52, McMaster-Carr, USA), and the feedstock chosen for verification is MetcoClad 316L-Si stainless steel powder (\#1079454, Oerlikon Metco AG, Switzerland; 45--106~\si{\micro\meter} particle size distribution). Table~\ref{tab:DED_Parameters} summarizes the nominal DED processing parameters used in this study.

\begin{table}[H]
    \centering
    \caption{DED printing parameters used in the experiment.}
    \begin{tabular}{|c|c|} \hline 
         \textbf{Printing Parameter} & \textbf{Nominal Value} \\ \hline 
         Laser power (W) & 600 \\ \hline 
         Laser diameter (mm) & 2.22 \\ \hline 
         Powder feed rate (g/min) & 14 \\ \hline 
         Scan speed (mm/s) & 7 \\ \hline 
         Hatch spacing (mm) & 0.8 \\ \hline 
         Interlayer step (mm) & 0.55 \\ \hline
    \end{tabular}
    \label{tab:DED_Parameters}
\end{table}

\subsection{Layer-Wise Print-Measure Loop}

\begin{figure}[H]
    \centering
    \includegraphics[width=0.95\linewidth]{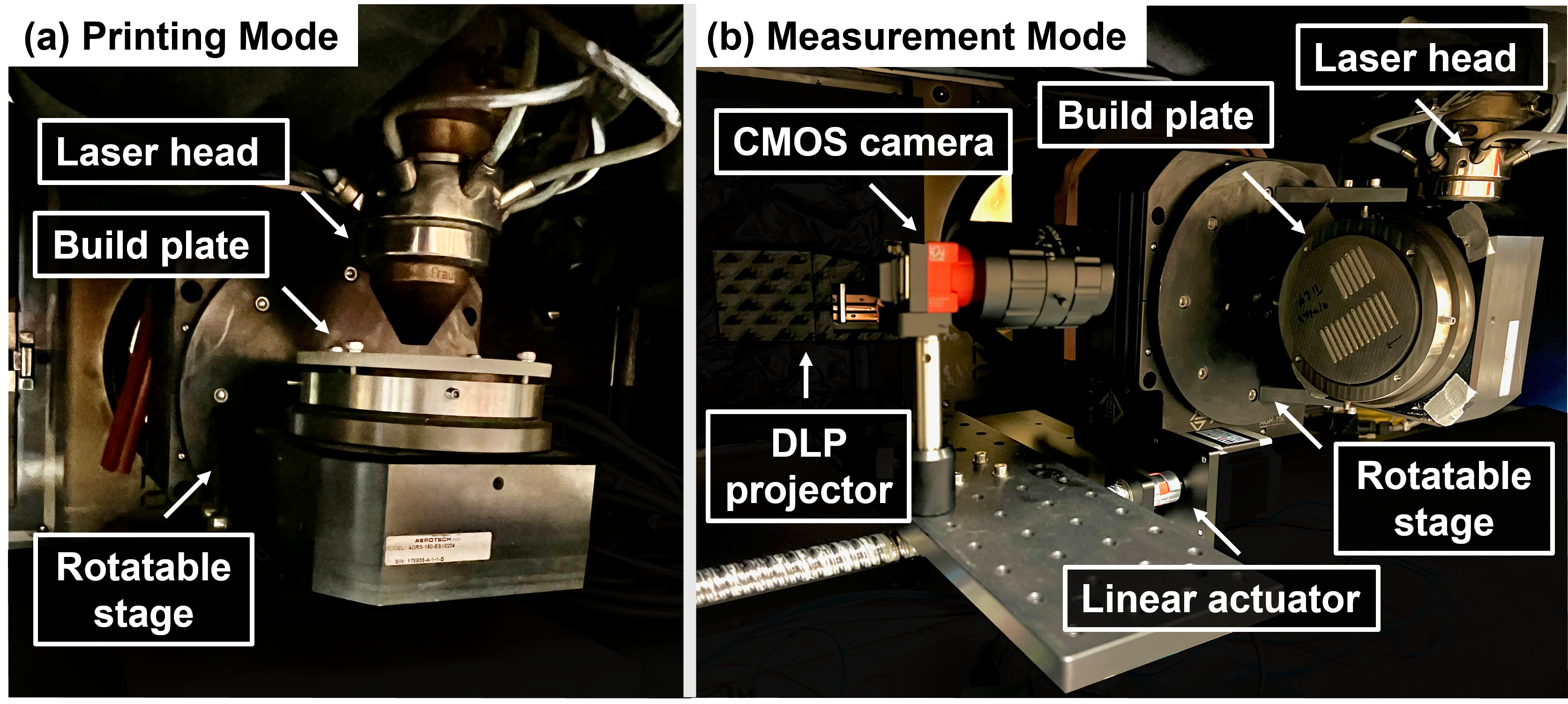}
    \caption{Integrated FPP setup within a DED system under two platform configurations: (a) printing mode and (b) measurement mode.}
    \label{fig:fringe_projection_system_layout}
\end{figure}

To facilitate interlayer surface monitoring during the DED process, a layer-wise print-measure loop was implemented, in which each deposited layer was immediately followed by an in-situ geometric measurement. This loop ensured that the surface morphology of each layer was recorded before the subsequent deposition, enabling timely evaluation of surface evolution and the early detection of potential anomalies. For each layer $n$, the system executed the following steps:

\begin{enumerate}
    \item \textbf{Printing}: The layer was deposited vertically using the powder-fed laser nozzle in the standard build orientation, as shown in Figure~\ref{fig:fringe_projection_system_layout}(a). 
    \item \textbf{Rotation}: Upon completion of the layer, the rotatable stage oriented the freshly printed surface toward the horizontally aligned FPP module, as shown in Figure~\ref{fig:fringe_projection_system_layout}(b). The substrate was rotated using an Aerotech AGR150 rotary stage with a direct encoder. This stage provides a bidirectional repeatability of \SI{39}{\micro rad}. Given the substrate diameter of $100~\mathrm{mm}$, the corresponding linear positioning error at the substrate edge (worst-case location) is \SI{\pm 1.95}{\micro\metre}. Since this value is less than $4\%$ of the system’s measurement accuracy (\SI{\pm 46}{\micro\metre}), the influence of rotation-induced positioning variation on the phase-to-height calibration can be neglected.

    \item \textbf{Measurement}: The FPP system captured fringe pattern images and reconstructed the surface topography via structured light projection.
    \item \textbf{Reorientation}: After imaging, the build plate was returned  to its original position for the deposition of the next.
\end{enumerate}

\begin{figure}[H]
    \centering
    \includegraphics[width=0.95\linewidth]{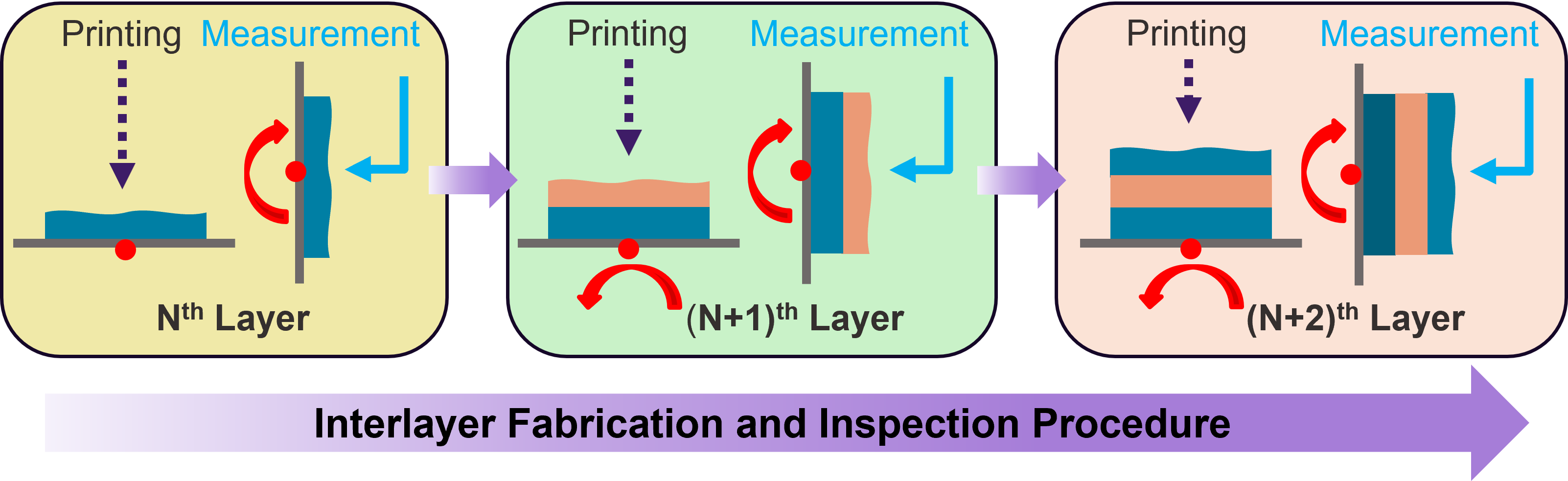}
    \caption{Schematic of the layer-wise print-measure loop. Each deposited layer is followed by a rotation of the build plate for in-situ FPP measurement before the next layer is printed.}
    \label{fig:print_measure_loop}
\end{figure}

This print-measure cycle was repeated throughout the build, as demonstrated for the $n^{\text{th}}$, $(n+1)^{\text{th}}$, and $(n+2)^{\text{th}}$ layers in Figure~\ref{fig:print_measure_loop}. A detailed timing breakdown for each cycle is provided in Table~\ref{tab:timing_breakdown}. During each build cycle, substrate flipping took approximately 1~s, and image acquisition required less than 2~s, resulting in a brief pause in the printing process. Considering that the deposition of a $12 \times 12~\mathrm{mm}$ layer required about 29~s, this short interruption ($<3~\mathrm{s}$) had a negligible effect on the overall build throughput. The subsequent phase calculation, 3D reconstruction, and anomaly detection were performed concurrently with the deposition of the next layer, further minimizing the measurement-induced impact on the overall process efficiency.

\begin{table}[H]
\centering
\footnotesize
\begin{threeparttable}
\caption{Timing breakdown of the measurement and processing cycle}
\label{tab:timing_breakdown}
\begin{tabular}{
>{\raggedright\arraybackslash}p{3.1cm}
>{\raggedright\arraybackslash}p{4.9cm}
>{\raggedright\arraybackslash}p{1.6cm}
>{\raggedright\arraybackslash}p{4.4cm}
}
\toprule
\textbf{Process step} & \textbf{Description} & \textbf{Time (s)} & \textbf{Notes} \\
\midrule
Substrate flipping and returning & Substrate rotation between measurement and deposition & $<$ 1 & Performed after each layer \\
\addlinespace
Image acquisition & Capture of 24 projected fringe patterns & $<$ 2 & Performed after each layer \\
\addlinespace
Phase calculation and 3D reconstruction & Processing one part area ($\sim$18 $\times$ 18 mm$^2$) & $\sim$ 3 & Executed concurrently during next-layer deposition \\
\addlinespace
NCR-based anomaly detection & Computation and identification using $>$2.2 million points & $\sim$ 5 & Executed concurrently during next-layer deposition \\
\midrule
\textbf{Total time per cycle} & \textbf{-} & \textbf{$<$ 11} & \textbf{Only $<$ 3 s causes deposition pause} \\
\bottomrule
\end{tabular}
% \begin{tablenotes}
% \scriptsize
% \item[a] ...
% \end{tablenotes}
\end{threeparttable}
\end{table}

This interleaved print-measure strategy enabled layer-by-layer full-field 3D surface acquisition at each layer, supporting cumulative assessment of part morphology and geometric consistency across layers. By documenting the morphological progression throughout the build, this loop established a basis for automated anomaly tracking and process control in future implementations.

\subsection{Design of Experiments}

To facilitate a systematic and repeatable evaluation of the FPP system’s performance in detecting surface irregularities, a series of controlled experiments was designed in which artificial geometric discontinuities were deliberately introduced. 

Prior research has shown that reducing laser power or powder feed rate can lead to insufficient layer height and degraded surface quality~\cite{hu_digital_nodate}; however, since laser power can be modulated almost instantaneously, under-deposition anomalies were introduced by momentarily shutting off the laser over short toolpath segments (1 mm) while keeping the powder flow constant. These interruptions provided a simplified and repeatable means of modeling surface anomalies. Due to melt pool dynamics and re-solidification behavior, the resulting surface geometry deviated from the idealized profile, enabling the investigation of how transient thermal disturbances affected the final morphology.

Based on this strategy, two anomaly distributions were fabricated:
(1) Six-anomaly set (one anomaly randomly inserted every other scan line).
(2) Twelve-anomaly set (two anomalies randomly inserted every other scan line).
A bilinear hatch infill strategy was applied over a $12 \times 12~\text{mm}^2$ area, with scan directions rotated by $90^\circ$ between successive layers to mitigate directional bias. All anomaly-containing layers were deposited atop previously built nominal layers to provide a stable and representative thermal environment during anomaly formation. To isolate the thermal effects of individual anomalies, each was spaced at a distance exceeding twice its length, thereby preventing melt pool overlap and ensuring independent thermal evolution.

\begin{figure}[H]
    \centering
    \includegraphics[width=0.95\linewidth]{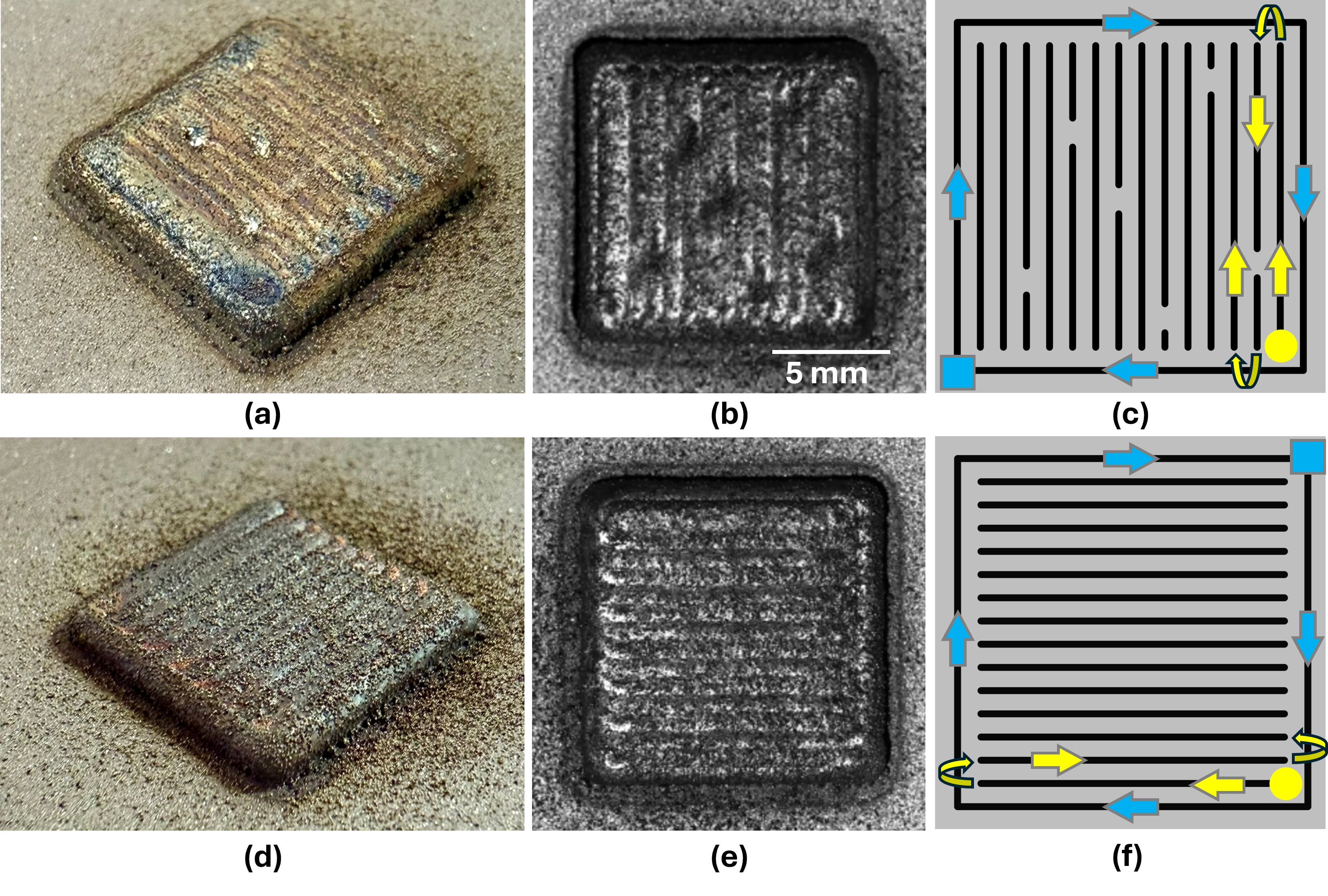}
    \caption{DED-fabricated surfaces with induced anomalies (a--c) and under nominal conditions (d--f): (a) and (d) post-build photographs; (b) and (e) in-situ grayscale images captured by the CMOS camera; and (c) and (f) laser scan toolpaths, with 1 mm laser-off segments indicating induced anomalies in (f).}
    \label{fig:Surface_optical_vs_FPP}
\end{figure}

A comparative visualization of parts fabricated with and without induced defects is illustrated in Figures~\ref{fig:Surface_optical_vs_FPP}. The post-build photographs captured with a digital single-lens reflex (DSLR) camera are shown in Figures~\ref{fig:Surface_optical_vs_FPP}(a) and \ref{fig:Surface_optical_vs_FPP}(d), while those in Figures~\ref{fig:Surface_optical_vs_FPP}(b) and \ref{fig:Surface_optical_vs_FPP}(e) present in-situ grayscale captures from the FPP system’s CMOS camera.

The designed laser scan paths corresponding to each build are illustrated in Figure~\ref{fig:Surface_optical_vs_FPP}(c) and \ref{fig:Surface_optical_vs_FPP}(f). In both cases, the printing process started at the corner indicated by the square marker, where the nozzle traces the outer boundary along the arrows to form a wall-like perimeter. It then proceeded from the circular marker, following the arrows to complete the inner hatch infill. The black lines indicate the programmed scan tracks. Interruptions along these lines represent 1 mm laser-off segments intentionally inserted to create defects, as shown in Figure~\ref{fig:Surface_optical_vs_FPP}(f). The 6- and 12-anomaly sets described above were primarily designed for a systematic evaluation of the FPP system's sensitivity to controlled, localized defects at complex, pre-determined toolpath locations. With these experimental conditions established, the subsequent analysis relies on the fringe projection system to capture the surface geometry after each deposited layer. Section~\ref{Anomaly_Detection_Framework} details the point cloud processing workflow and demonstrates how anomalies can be identified and characterized based on the measured data. 

However, to validate the system's generalizability and effectiveness in a more ecologically valid scenario, a second experiment was conducted. This experiment was designed to replicate a common process failure mode. We fabricated an entire layer using only 50\% of the nominal laser power. This condition is known to produce realistic LOF morphologies and widespread surface degradation, which are often encountered before process parameters are fully optimized.

The purpose of this test was to demonstrate that the anomaly detection framework can generalize from the artificial, discrete defects (i.e., the laser-off segments) to these process-induced, stochastic defects, thereby proving its utility on surface morphologies representative of a true manufacturing flaw. This 50\%-power layer, along with a baseline surface deposited under nominal conditions, was analyzed using the identical FPP workflow.

\section{Anomaly Detection Framework}
\label{Anomaly_Detection_Framework}

\subsection{Point Cloud Processing}
\label{Point_Cloud_Processing}

To accurately characterize the surface morphology of DED-fabricated parts, the raw phase data acquired from the FPP system were converted into 3D point clouds. While depth maps offer a 2D visualization of surface elevation, they are limited in capturing fine-scale geometric details. In contrast, 3D point clouds provide a richer spatial representation, enabling both visual assessment and quantitative analysis through methods such as curvature estimation, surface normal computation, and localized anomaly segmentation.

\begin{figure}[H]
    \centering
    \includegraphics[width=0.48\linewidth]{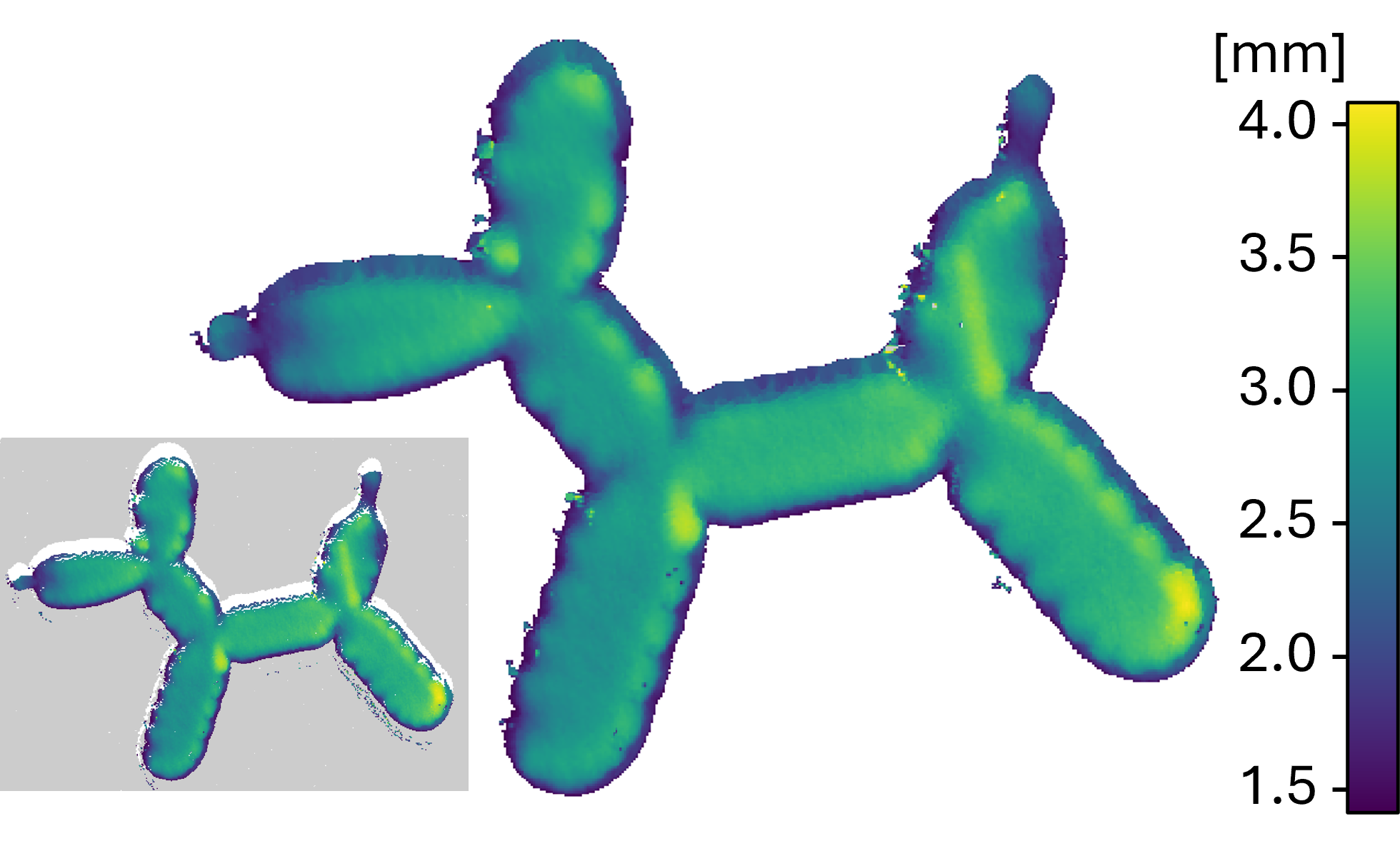}
    \caption{Measurement results for an irregular DED-printed specimen (inset) and top-surface extraction using Otsu’s thresholding.}
    \label{fig:Balloon_dog_white}
\end{figure}

Phase data was first translated into physical height using a pre-calibrated phase-to-height conversion constant $C$, allowing unwrapped phase values to be directly mapped to vertical coordinates. The resulting depth map captured surface variation across both the deposited material and the surrounding build plate, with lateral resolution defined in pixel units. To isolate the printed geometry and eliminate background data, Otsu’s thresholding method was applied to segment the depth map based on height intensity \cite{chen_automatic_2021}. This segmentation preserved boundary integrity, ensuring accurate extraction of the deposited area for subsequent analysis. The primary objective of this step is to accurately identify the region of interest (ROI) corresponding to the newly deposited layer. For complex geometries, the ROI can be further localized based on the programmed deposition path, which provides prior knowledge of the target area. Because the measurement consistently targets the top surface after each deposition, challenges such as line-of-sight occlusion or large surface variations have negligible influence on the surface measurement of the newly deposited layer. Figure~\ref{fig:Balloon_dog_white} shows a processing example for an irregular DED-printed specimen with a non-planar and more complex geometry. The inset presents the full measured morphology, while the main figure displays the ROI obtained after applying Otsu's thresholding method. This example demonstrates that the same processing pipeline is applicable to more complex geometries.

After the ROI was extracted for the primary specimen, the segmented region was transformed into a 3D point cloud, where lateral coordinates were computed using the pixel-to-length mapping obtained during intrinsic camera calibration, yielding an effective spatial resolution of approximately $\SI{12}{\micro\meter} \times \SI{12}{\micro\meter}$.

\begin{figure}[H]
    \centering
    \includegraphics[width=0.48\linewidth]{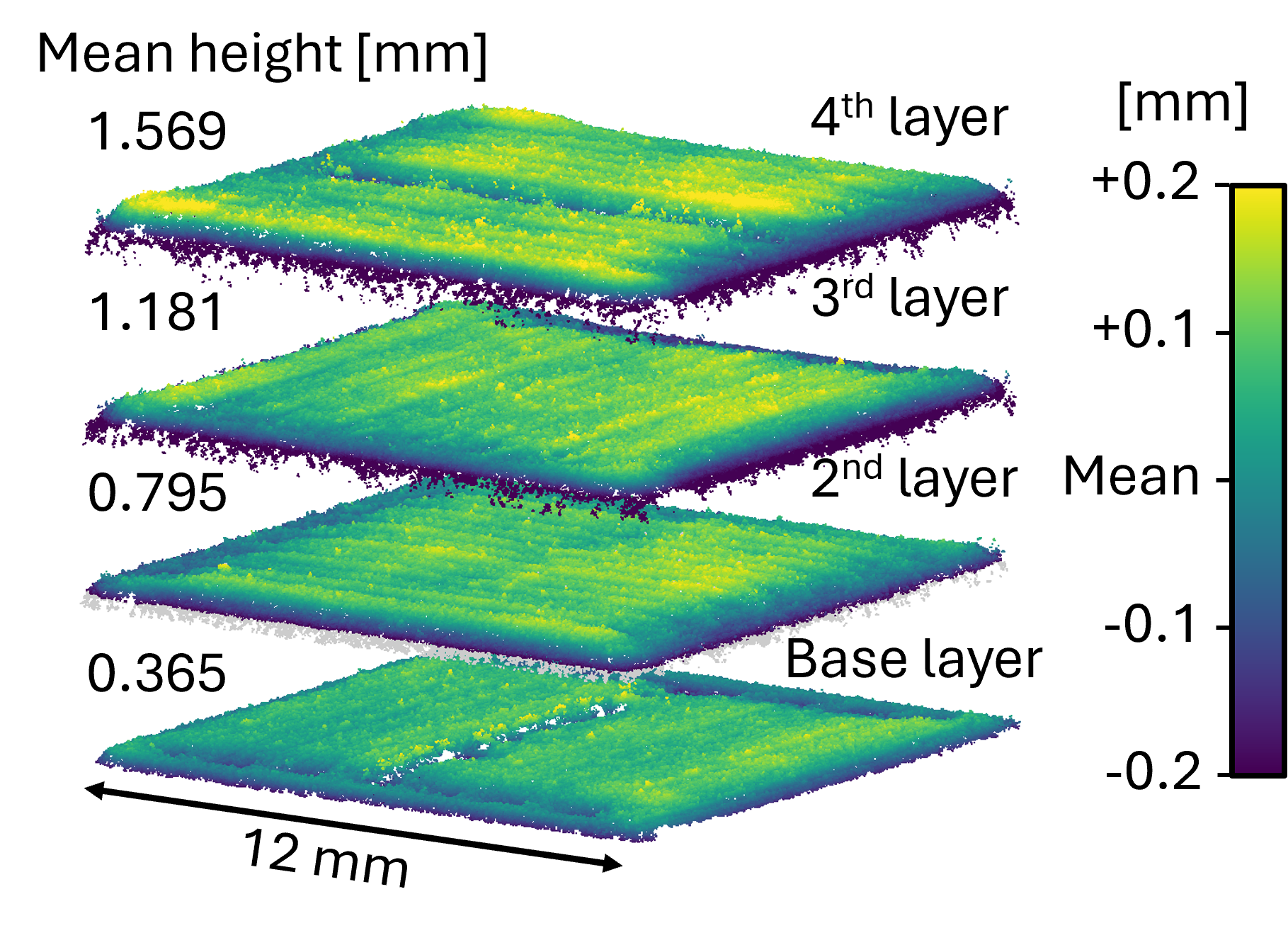}
    \caption{Sequential relative height maps of a single DED-printed part acquired via FPP after each deposition step, with height values expressed as deviations from the mean height of the corresponding layer.}
    \label{fig:Layer_sequences}
\end{figure}

Despite accurate calibration, the raw point clouds often contain noise and outliers caused by optical reflections, measurement artifacts, or partial occlusions. To mitigate these effects, a statistical outlier removal (SOR) filter was applied~\cite{rusu_3d_2011}. In this approach, for each point in the cloud, the mean Euclidean distance to its $k = 6$ nearest neighbors was computed. A global analysis of these distances across the dataset yielded the mean $\bar{d}$ and standard deviation $\sigma$. Points whose neighborhood distance exceeded a defined threshold were classified as outliers and removed. The threshold was calculated as:
\begin{equation}
d_{\text{max}} = \bar{d} + \lambda \cdot \sigma,
\end{equation}
where $\lambda = 1$ is the standard deviation multiplier. Any point with a mean neighborhood distance $d_i > d_{\text{max}}$ was discarded as a statistical outlier. After applying the SOR filter, the denoised point clouds were reconstructed for each deposition step, enabling a layer-by-layer visualization of the evolving surface morphology of the printed part, as presented in Figure~\ref{fig:Layer_sequences}.

\begin{figure}[H]
    \centering
    \includegraphics[width=0.48\linewidth]{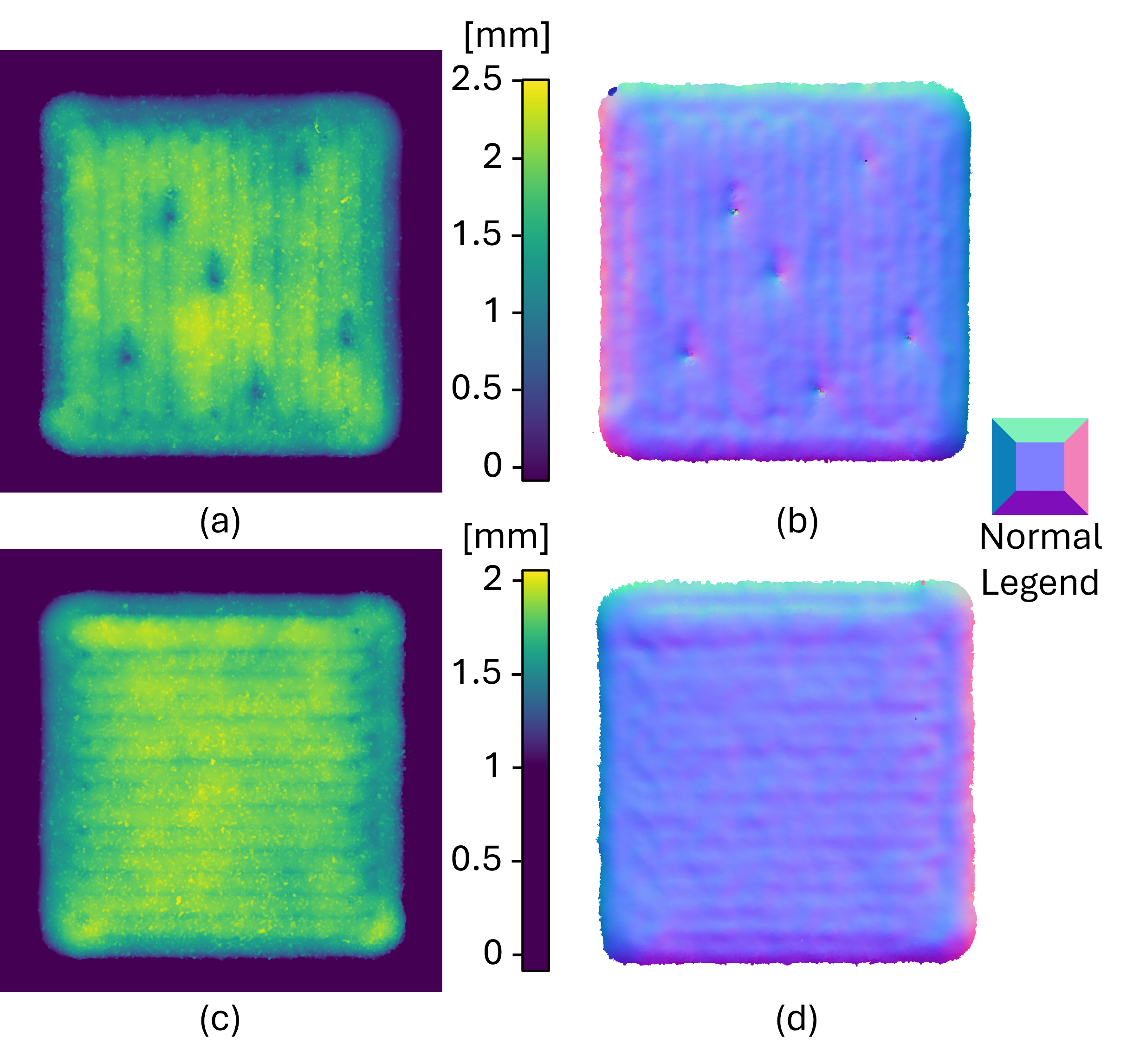}
    \caption{Surface measurements obtained using FPP. (a) and (b) correspond to a part with intentionally induced anomalies, while (c) and (d) correspond to a part fabricated under nominal conditions. For both cases, (a) and (c) present the reconstructed height maps, and (b) and (d) show the corresponding surface normal maps.}
    \label{fig:Height_and_normal_maps}
\end{figure}

Surface reconstruction results based on both height and surface normal maps are shown in Figure~\ref{fig:Height_and_normal_maps}. In the defective sample, the height map reveals localized depressions and irregularities (Figure~\ref{fig:Height_and_normal_maps}(a)), while the corresponding normal map highlights sharp angular transitions and curvature variations (Figure~\ref{fig:Height_and_normal_maps}(b)), indicating geometric discontinuities caused by the induced anomalies. By contrast, the nominal sample (Figure~\ref{fig:Height_and_normal_maps}(c) and \ref{fig:Height_and_normal_maps}(d)) exhibits a smooth surface profile with minimal angular deviation, confirming consistent deposition quality. To quantitatively assess surface irregularities in DED-fabricated parts, two geometric descriptors were extracted from the reconstructed point clouds: local point density and the normal-change rate (NCR) method. These metrics characterize surface morphology at different scales, enabling the detection of both fine texture variations and larger geometric anomalies.

\subsection{Surface Anomaly Visualization and Segmentation}

The local point density was computed as the number of neighbors within a fixed-radius search, the search radius defines the spherical spatial extent surrounding each query point within which neighboring points are identified for metric calculation~\cite{zhou_open3d_2018}. Let $\mathcal{P} = \{p_1, p_2, \dots, p_n\}$ denote the set of 3D points sampled from the reconstructed surface. For each point $p_i \in \mathcal{P}$, the neighbors within a search radius $r$ were counted as: 

\begin{equation}
N(p_i) = \left| \left\{ p_j \in \mathcal{P} \mid \|p_j - p_i\| \leq r \right\} \right|.
\end{equation}

\added{In this study, the search radius $r$ was set to 0.3 mm, a value specifically chosen to be approximately twice the characteristic size of the target powder residue anomalies (typically ranging from 0.15 mm to 0.3 mm). This choice represents a balance between sensitivity to localized geometric variations and robustness against intrinsic surface roughness and measurement noise inherent to as-built DED surfaces. While a smaller radius might capture finer textures, it would also elevate false-positive rates due to noise. Conversely, a larger radius would over-smooth local variations and reduce the detectability of small-scale surface anomalies. Furthermore, considering the system's lateral resolution of 12 \textmu m, a perfectly planar surface would theoretically yield approximately 1,960 neighboring points within this 0.3 mm radius. This chosen radius ensures a sufficient number of neighboring points are included to yield statistically meaningful local density estimates.} This metric captures small-scale features such as partially sintered powder residues and local undulations between deposition paths. \added{Regions with low neighbor counts typically correspond to voids or occluded areas arising from steep surface curvatures or aggressive noise filtering during post-processing. In the context of DED monitoring, such data sparsity constitutes an informative anomaly signature rather than a mere measurement artifact, as steep occlusions are generally associated with sharp geometric transitions or deep surface voids. Furthermore, missing points caused by specular reflections, which are subsequently removed by the SOR filter, are observed to be sparse and spatially dispersed in our experiments. Unlike the clustered missing-point patterns induced by genuine defects, these isolated data gaps do not accumulate sufficiently within the chosen search radius to exceed the anomaly detection thresholds.} Conversely, high neighbor counts indicate regions of material accumulation or smooth, overbuilt surfaces.

\begin{figure}[H]
    \centering
    \includegraphics[width=0.95\linewidth]{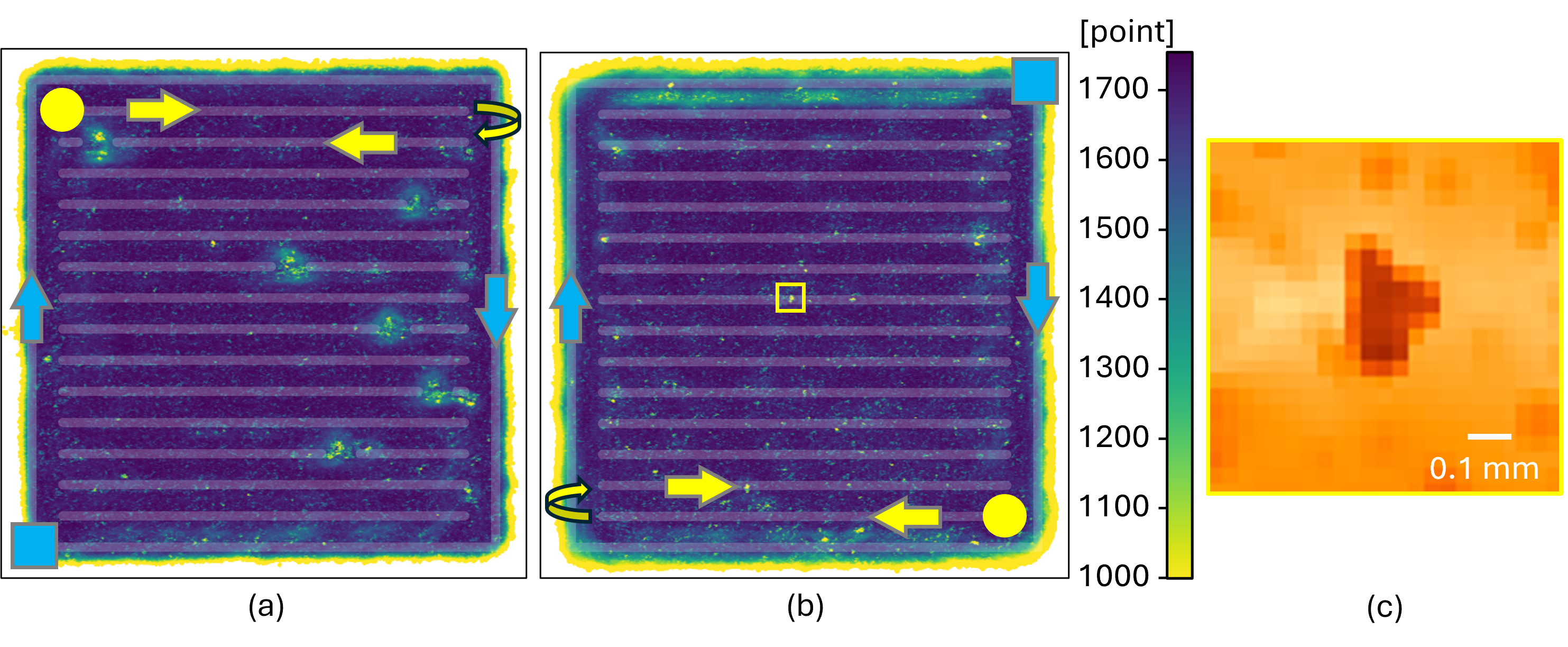}
    \caption{Local point cloud density of DED-fabricated parts with a neighborhood radius of 0.3 mm. (a) Part with intentional anomalies; (b) part built with nominal parameters. (c) 3D microscopy ($5\times$) zoom-in of the outlined region in (b).}
    \label{fig:Cloud_density}
\end{figure}

The surface, containing six laser-off anomalies, is illustrated in Figure~\ref{fig:Cloud_density}(a). Disruptions in the programmed toolpath produce discontinuities in deposition, yielding elevated neighbor densities at melt pool termination sites and geometrically irregular regions of incomplete fusion. In contrast, Figure~\ref{fig:Cloud_density}(b) shows overall uniform density, though sporadic particulates ($\sim$150~\textmu m) remain detectable. Sparse neighbor counts are consistently observed near hatch turnarounds, reflecting susceptibility to LOF voids due to reduced thermal continuity during path reversal. Additionally, a prominent discontinuity between raster fill and the pre-existing wall highlights the impact of deposition timing: the wall solidified prior to adjacent infill, preventing sufficient consolidation and resulting in a low-density gap. The validity of the density-based metric is confirmed in Figure~\ref{fig:Cloud_density}(c), where a focus-variation scan of the highlighted region resolves a representative surface particle (197~\textmu m $\times$ 140~\textmu m lateral dimensions, $\sim$183~\textmu m in height). This agreement demonstrates that the neighbor-count approach reliably reflects micro-scale roughness features and provides an intuitive mapping of DED process-induced anomalies.

Surface curvature was characterized using the NCR method~\cite{rusu_3d_2011}. For each point $p_i$, a normal vector $\mathbf{n}_i$ was estimated within a localized neighborhood. \added{The NCR search radius was calibrated to 0.6 mm, which is comparable to the LOF voids in this study (typically 0.8--1.2 mm) Unlike powder residues, LOF defects are geometrically distinct from the background roughness and possess a larger topographical footprint. Consequently, their geometry encompasses a sufficient number of data points to yield a stable curvature signal without the need for the expanded filtering-to-feature ratio required for smaller anomalies.} For every neighboring point $p_j \in N(p_i)$, the corresponding normal $\mathbf{n}_j$ was also computed. The NCR value at $p_i$ is defined as the mean angular deviation between its own normal and the normals of all neighbors: 

\begin{equation}
\text{NCR}(p_i) = \frac{1}{|N(p_i)|} \sum_{p_j \in N(p_i)} \arccos\left( \left| \mathbf{n}_i \cdot \mathbf{n}_j \right| \right).
\end{equation}

\begin{figure}[H]
    \centering
    \includegraphics[width=0.95\linewidth]{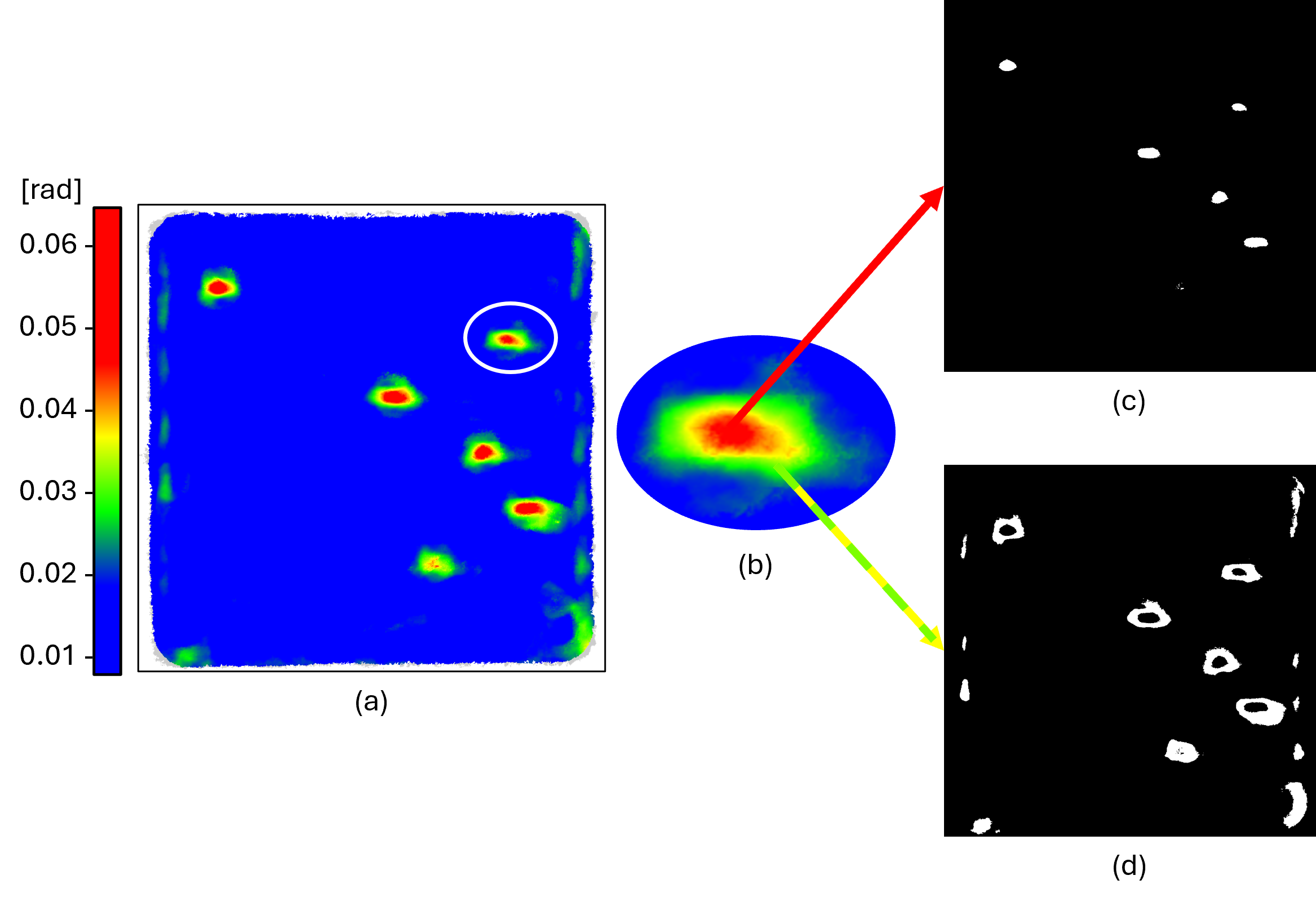}
    \caption{Anomaly segmentation using the NCR metric. (a) NCR map with high values in red; (b) zoomed anomaly; (c) binary segmentation of high-NCR cores; (d) segmentation of medium-NCR transitional regions.}
    \label{fig:Defect_segmentation_pipeline}
\end{figure}

Here, $i$ identifies the reference point and $j$ indexes its neighbors. The absolute value ensures that normals are compared regardless of orientation. High NCR values indicate abrupt curvature changes from under-/over-deposition or geometric irregularities. NCR maps were visualized using sequential colormaps, where red zones highlight anomalies such as conical depressions (Figure~\ref{fig:Defect_segmentation_pipeline}). 

A multi-threshold segmentation strategy was then applied to different severity. \added{In this study, an NCR value greater than 0.045 rad was defined as a high-NCR threshold to isolate the cores of severe discontinuities, while a threshold of 0.018 rad was used to capture the surrounding medium-NCR transitional regions.}  This manual approach is a practical choice, as it allows researchers to define ``defect'' severity based on process-specific criteria, such as distinguishing critical, propagating flaws or minor, acceptable roughness. This enables hierarchical segmentation of anomalies and their geometric extent.

\begin{figure}[H]
    \centering
    \includegraphics[width=0.95\linewidth]{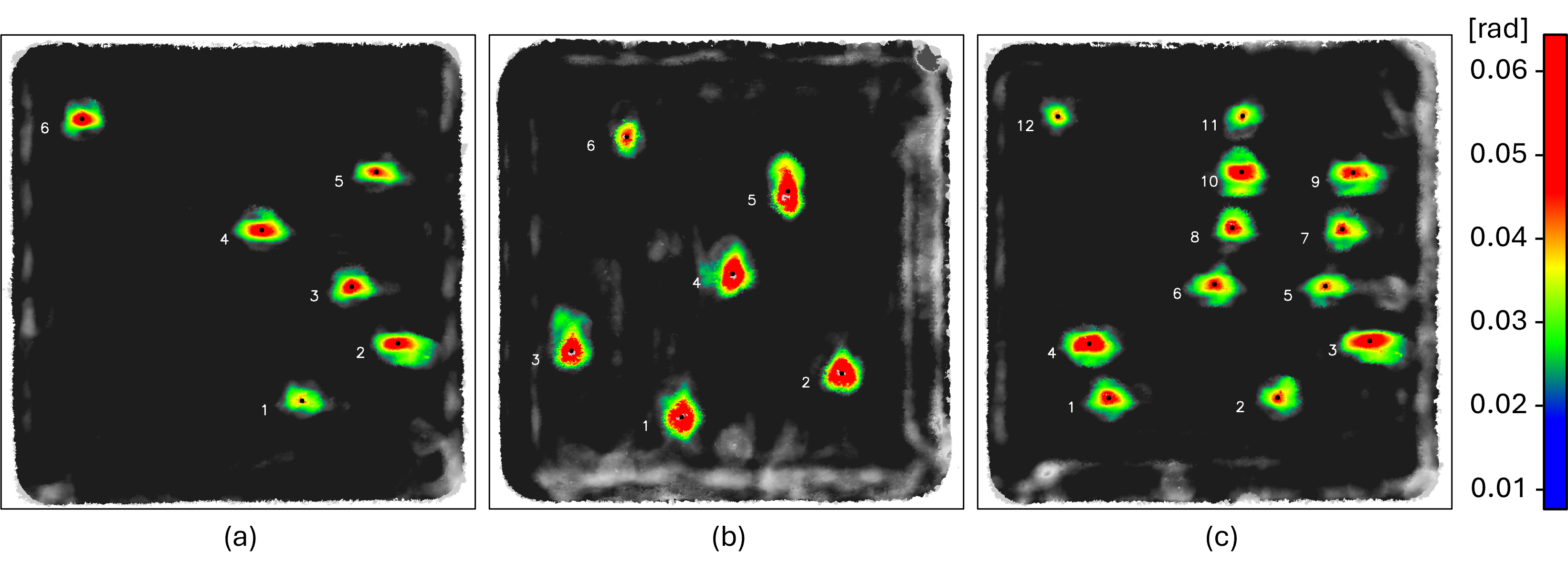}
    \caption{NCR-based anomaly maps. (a) Six intentional laser-off anomalies; (b) six laser-off anomalies with vertical inner hatch infill; and (c) a sample with twelve anomalies. Red zones denote high curvature variation, and black dots mark anomaly centroids.}
    \label{fig:Defect_index_maps}
\end{figure}

The proposed framework leverages complementary descriptors: point density reflects spatial completeness, whereas NCR emphasizes curvature transitions. Figure~\ref{fig:Defect_index_maps} shows the segmented anomalies across samples with six or twelve laser-off defects. The elongated, elliptical morphologies align with scan directions, reflecting melt pool dynamics under transient beam interruptions. 

To validate detection performance, the segmented anomaly centroids were cross-referenced with the pre-defined coordinates of the laser interruptions. With thresholds empirically calibrated to the characteristic scale of the defects, the system yielded a 100\% detection rate for all designed anomalies. Furthermore, visual verification against the corresponding high-contrast depth maps confirmed that the NCR-defined boundaries accurately capture the full topographical extent of the depressions. It should be noted that these results represent performance under optimized thresholding. In a practical industrial scenario, the selection of thresholds involves a trade-off between sensitivity and specificity. Lowering the thresholds to detect subtler, smaller-scale defects would likely increase the false-alarm rate due to surface roughness noise. Conversely, stricter thresholds minimize false alarms but may miss marginal anomalies. Therefore, the 100\% detection rate reported here demonstrates the system's capability to isolate significant geometric deviations when parameters are appropriately tuned.

In addition to the intentionally designed defects, the bright gray regions on the surface also correspond to areas with higher NCR values, indicating uneven topography. These regions, particularly along the outer edges, are attributed to insufficient lateral support at the melt-pool boundary, leading to gradual geometric drift as the build height increases. Although such macroscopic deviations were not identified as defects in the current segmentation results because their NCR values did not exceed the defined high-threshold criterion, they represent another important class of process-induced anomalies that can be quantitatively evaluated in future studies using the same geometry-based framework.

\begin{figure}[H]
    \centering
    \includegraphics[width=0.48\linewidth]{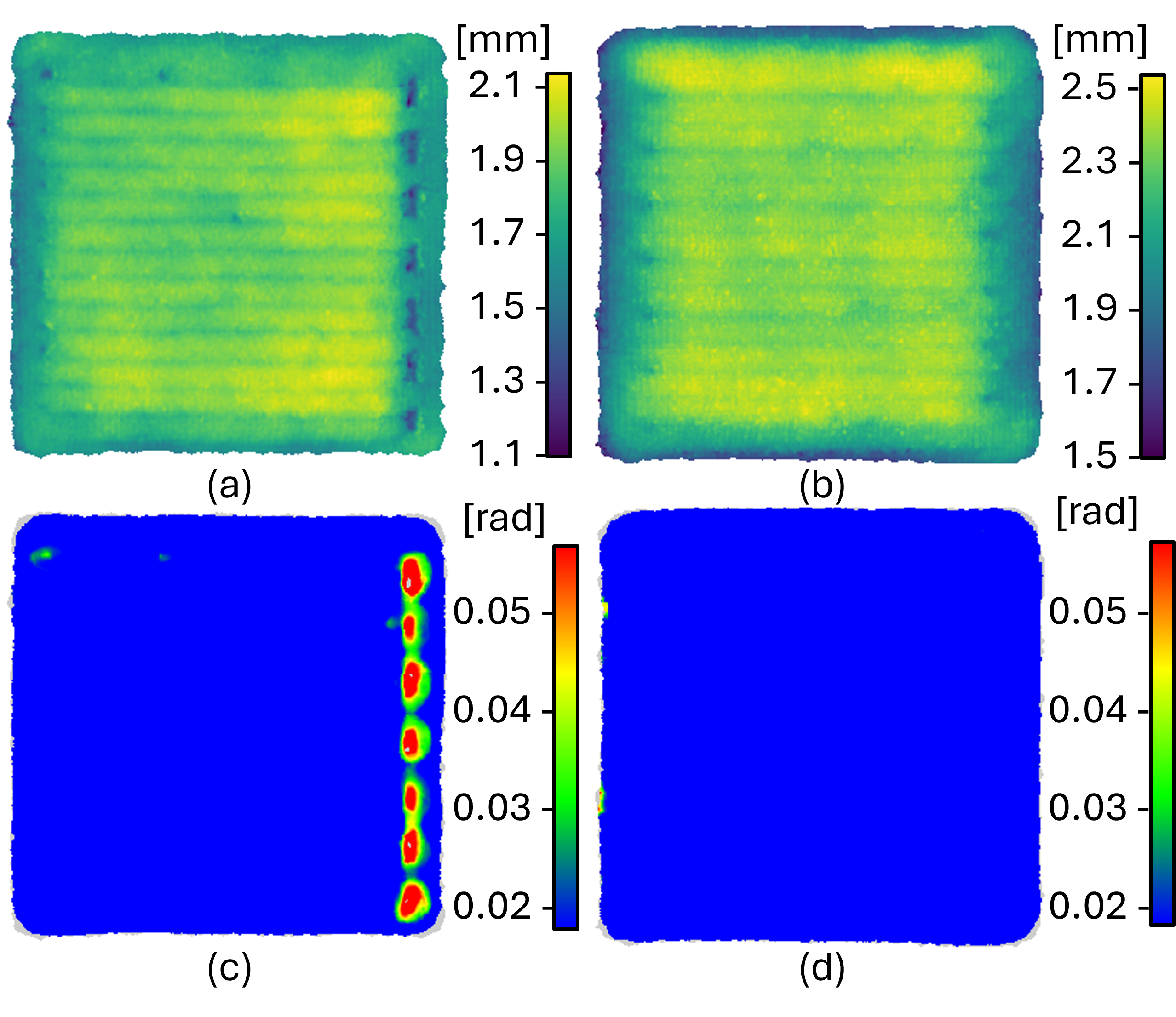}
    \caption{Validation on process-induced flaws comparing a 50\% power sample (left) and a nominal sample (right). (a) Depth maps show periodic pits in the 50\% sample. NCR maps confirm selective segmentation of these pits (c) while showing no anomalies on the nominal surface (d).}
    \label{fig:Real_defect_norminal_white}
\end{figure}

To validate the framework's generalizability on realistic, process-induced flaws, samples fabricated under nominal conditions and with a 50\% reduction in laser power were analyzed. Figure~\ref{fig:Real_defect_norminal_white} compares these two conditions, displaying the depth map and the corresponding NCR-based segmentation maps for the 50\%-power sample and the nominal sample.
The depth map of the 50\%-power sample (Figure~\ref{fig:Real_defect_norminal_white}(a)) exhibits periodic pits at the toolpath turning points, visible along the right edge. These ``lack of fusion'' type anomalies are absent in the nominal sample (Figure~\ref{fig:Real_defect_norminal_white}(b)). The corresponding NCR map (Figure~\ref{fig:Real_defect_norminal_white}(c)) shows that, by adjusting the segmentation threshold, the system segmented these critical pits while registering no false positives in the surrounding high-NCR areas. In contrast, the nominal sample (Figure~\ref{fig:Real_defect_norminal_white}(d)) shows no significant NCR-flagged regions, confirming the method's specificity.

This comparison highlights the framework's selective segmentation. While the depth map of the 50\%-power sample shows other minor, stochastic surface variations, the NCR thresholding filters out these regions of less severe topographical change. This selectivity allows for differentiation based on defect severity, focusing on significant geometric flaws (like the turning point pits) while disregarding minor, acceptable roughness.

\added{Notably, all anomaly detection results presented were obtained using an identical set of hyperparameter, without retuning for different samples. This consistency suggests that the proposed geometry-based descriptors are relatively insensitive to surface appearance variations and exhibit promising robustness across different DED surface morphologies within the studied parameter space.}

\subsection{Comparison to 2D Image Processing Methods}

\begin{figure}[H]
    \centering
    \includegraphics[width=0.95\linewidth]{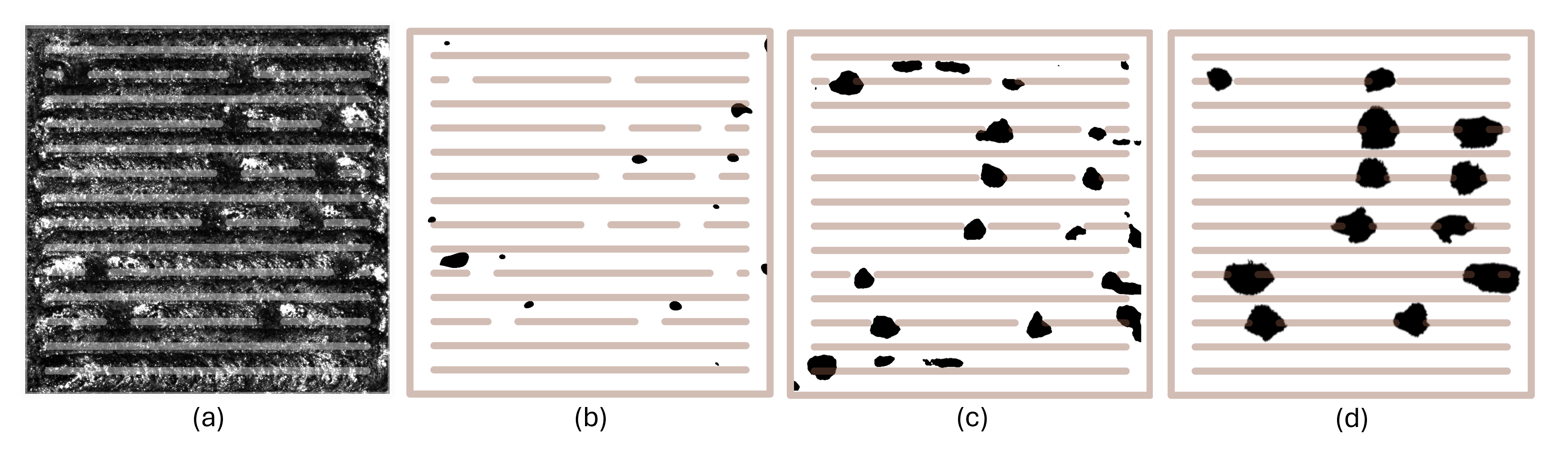}
    \caption{Comparison of anomaly highlighting in DED surface using 2D image-based and 3D point-cloud-based methods. (a) Input image with twelve anomalies; (b) global Yen thresholding (2D); (c) local Sauvola thresholding (2D); and (d) NCR-based 3D point cloud analysis.}
    \label{fig:Benchmark}
\end{figure}

To provide a comparison with conventional techniques (non-learning-based), several image processing methods that are commonly employed in the literature for preprocessing and anomaly highlighting in 2D PBF image analysis were applied to DED surface data~\cite{li_simultaneous_2024, dinh_layering_2021}. As illustrated in Figure~\ref{fig:Benchmark}(a), a surface containing twelve defects was reconstructed by averaging FPP images over one fringe-shifting cycle, followed by two iterations of contrast enhancement at a saturation parameter of 0.35, which served as the input image subsequently processed with a Gaussian blur ($\sigma = 10$). The result obtained using the Yen global thresholding method (Figure~\ref{fig:Benchmark}(b)) predominantly produced large, isolated overexposed regions; however, these regions did not permit precise localization of individual anomalies and led to a high false-positive rate. In contrast, local Sauvola thresholding with a window size of 100 (Figure~\ref{fig:Benchmark}(c)) captured most true anomalies but frequently misclassified benign features such as scan path overlaps and turnaround grooves. Although post-processing using geometric descriptors, such as area, circularity, and aspect ratio, alleviated some of these misclassifications, the performance remained highly sensitive to threshold selection and illumination variations, requiring frequent manual tuning to sustain acceptable accuracy. 

Even though classical algorithms like Yen and Sauvola do not represent the current state-of-the-art in 2D inspection, which increasingly relies on machine learning (ML) techniques. However, this comparison serves to illustrate a fundamental challenge inherent to any method relying solely on 2D intensity data, the confounding of true surface geometry with imaging artifacts (e.g., specular highlights, shadows, and material reflectivity).

As shown, the 2D methods mistook these high-contrast, non-geometric features for defects. An advanced 2D ML model, like an unsupervised autoencoder trained on this same flawed intensity data, would still be susceptible to these artifacts. It would likely learn to flag these high-contrast regions, leading to false positives, or require extensive and complex training data to learn to ignore them, which is a task made trivial by using 3D data.

In contrast, NCR analysis of 3D point clouds demonstrated superior robustness for anomaly segmentation in DED (Figure~\ref{fig:Benchmark}(d)). Unlike 2D threshold-based methods adapted from image preprocessing in PBF studies, NCR leverages proportional geometric thresholds in 3D space, making it inherently less affected by surface brightness fluctuations and imaging artifacts. With a lateral resolution of approximately $\sim$12~\textmu m/pixel, the NCR-based system enabled reliable quantitative extraction of defect-related descriptors, including anomaly count, spatial distribution, centroid deviation, and morphological metrics. This capability not only improved segmentation accuracy but also established a scalable framework for process diagnostics and parameter optimization in laser-DED, thereby overcoming key limitations of conventional 2D image-based anomaly highlighting approaches.

\subsection{Characterization of Anomalies}

\begin{figure}[H]
    \centering
    \includegraphics[width=0.95\linewidth]{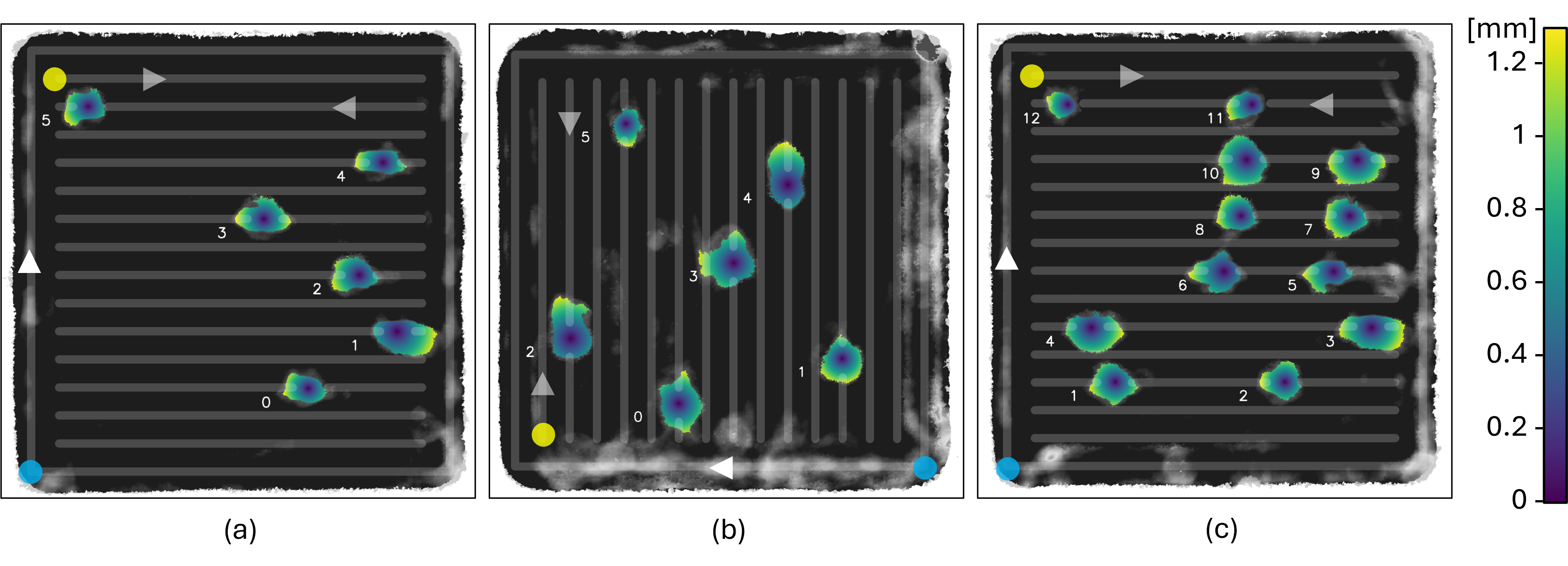}
    \caption{Pixel-wise distance mapping between programmed laser-off points and segmented surface anomalies in three DED-fabricated samples. The colormap encodes 2D Euclidean distance from each pixel within an anomaly region to the nearest laser-off center.}
    \label{fig:Defect_distance_to_laser_off}
\end{figure}

To assess the correspondence between programmed defect locations and observed surface morphology, a pixel-wise distance map was computed from segmented anomaly regions to the nearest laser-off center along the toolpath (Figure~\ref{fig:Defect_distance_to_laser_off}). The colormap encodes the Euclidean distance from each anomaly pixel to its respective shut-off point, thereby visualizing both spatial extent and directional bias relative to the scan trajectory.

A recurring feature is the emergence of elongated ``tails'' extending opposite to the scan direction, frequently represented as yellow streaks in the distance maps. This asymmetric morphology arises because, although the laser was deactivated instantaneously, powder feed and carrier gas delivery continued. The arrival of unmelted powder and cold gas onto a still-hot melt pool accelerated local cooling, inducing thermal shock and premature solidification that distorted the downstream anomaly geometry. This phenomenon highlights a thermal inertia effect, where the combined influence of residual melt pool heat and ongoing material deposition drives the formation of elongated morphological artifacts beyond the intended anomaly site. Quantifying the offset between the geometric centroid of the anomaly and the nominal laser-off point provides valuable insight into the dynamic coupling between heat flow, powder-gas interaction, and material solidification during transient deposition events.

\section{Conclusions}

This study demonstrates that a build-height–synchronized FPP module enables micrometer-scale, full-field 3D reconstruction of DED surfaces with a fidelity of $\pm\SI{46}{\micro\meter}$. The method ensures consistent temporal tracking of surface evolution and allows for the early detection of geometric deviations before they propagate into hidden defects. While the primary objective was to validate the system’s in-situ measurement capability and demonstrate a concept for localized surface anomaly detection, the proposed framework, which relies on point cloud density and normal phase change, encodes reliable signatures of deposition quality, reducing the reliance on manual annotation. Crucially, this framework also inherently captures large-scale geometric distortions, such as cumulative edge deformation, thereby shifting monitoring from indirect process proxies to direct geometric evidence and offering a new perspective on process stability.

Future work will focus on expanding this geometry-based framework to quantitatively evaluate the evolution of such large-scale deviations. The reconstructed 3D surface models will be compared with the original design geometry to track cumulative distortion. By setting tolerance thresholds for accumulated deviation, the system can support an automated monitoring strategy for both local and global geometric anomalies. \added{Furthermore, to fundamentally mitigate blind spots caused by steep occlusions, the system will be extended to multi-view FPP configurations. This approach will fuse data from multiple angles to ensure complete surface coverage, further enhancing robustness.} Such multimodal approaches will enable more comprehensive defect characterization and support the development of closed-loop process control strategies in DED.

\section*{Acknowledgment}
This research was supported by the U.S. National Science Foundation (NSF) under the Engineering Research Center for Hybrid Autonomous Manufacturing Moving from Evolution to Revolution (ERC-HAMMER, Award EEC-2133630), and under Grants CMMI-2216298 and DGE-2234667.

\appendix
% \section{Example Appendix Section}
\label{app1}

 \bibliographystyle{elsarticle-num} 
 \bibliography{references}



\end{document}